\documentclass[
superscriptaddress,
amsmath,amssymb,
aps,
pra,
a4paper,twocolumn,
]{revtex4-2}

\usepackage[european]{circuitikz} 
\usepackage{graphicx}
\usepackage{bm}
\usepackage{subcaption}
\usepackage{ragged2e}

\usepackage[english]{babel}
\usepackage{soul}
\usepackage[normalem]{ulem}
\usepackage{bbold}
\usepackage{braket}
\usepackage{xcolor}
\usepackage{csquotes}
\usepackage{appendix}
\usepackage{array}
\usepackage{hyperref}
\usepackage{float}
\hypersetup{
    colorlinks=true,
    linkcolor=blue,
    filecolor=magenta,      
    urlcolor=cyan,
    anchorcolor = black
    }

\usepackage{overpic}

\begin{document}


\newcommand{\bmgate}{\mathbf{U}^{\rm BM}}
\newcommand{\usin}{\mathcal{U}^{\rm sin}}
\newcommand{\ubm}{\mathcal{U}^{\rm BM}}
\newcommand{\plaquette}{\mathbf{U}^{\varkappa }}
\newcommand{\uscone}{\mathcal{U}_{\rm sc+1 }}
\newcommand{\usc}{\mathcal{U}_{\rm sc }}

\title{A nonlinear quantum neural network framework for entanglement engineering }

\author{Adriano Macarone-Palmieri}\email{adriano.macaronepalmieri@unipa.it}
\affiliation{Dipartimento di Ingegneria, Università degli Studi di Palermo, Viale delle Scienze, 90128 Palermo, Italy}

\author{Alberto Ferrara}
\affiliation{Dipartimento di Ingegneria, Università degli Studi di Palermo, Viale delle Scienze, 90128 Palermo, Italy}

\author{Rosario Lo Franco}
\affiliation{Dipartimento di Ingegneria, Università degli Studi di Palermo, Viale delle Scienze, 90128 Palermo, Italy}


\date{\today}

\begin{abstract}
Multipartite entanglement is a crucial resource for quantum technologies; however, its scalable generation in noisy quantum devices remains a significant challenge. Here, we propose a low-depth quantum neural network architecture with linear scaling, utilizing a novel approach to introducing activation functions for the task of entanglement engineering. 
As a testbed to demonstrate the clear advantage unlocked by the introduction of nonlinear activations, we run a Monte Carlo sampling over $10^5$ circuit topologies for pure noiseless states.
Subsequently, we focus our attention on the noisy scenario; we employ the experimentally accessible Meyer–Wallach global entanglement as a scalable surrogate optimization cost and certify entanglement using bipartite negativity. For 10-qubit mixed states, the optimized circuits generate substantial entanglement across the bipartitions. Lastly,  the presence of genuine multipartite entanglement is certified with semi-definitive programming. These results establish an experimentally motivated and scalable framework for engineering multipartite entanglement on near-term quantum devices, highlighting the combined role of nonlinearity and circuit topology scaling up to 20 qubits readily.
\end{abstract}

\maketitle

\section{Introduction}

Alongside quantum error correction, entanglement engineering is a cornerstone of next-generation quantum communication, sensing, and computation \cite{Review-QuantumEntaglement, Review_QuantumSensing,EntBasedTechnology}. Multipartite entangled states enable enhanced security, sensitivity, and coherence, yet their scalable preparation in realistic quantum devices remains a central challenge.

Entanglement engineering can be formally cast as a graph-optimization problem, in which both interaction structure and parametrization determine the entanglement properties of the resulting quantum state \cite{QuantumExperimentsAndGraphs, EntanglementGenerationKobra}. This perspective has motivated the use of machine-learning techniques, particularly optimization and reinforcement learning, in quantum information science \cite{Review_AIforQauntum-Krenn, ReviewAIforQuantumComp-2025,RL_subgraphOptimization}. Such approaches have been successfully applied to the automated design of quantum experiments \cite{VAE_EntanglementExperiments-2022}, circuit synthesis \cite{RL4qEntangledQuibits, DiffusionCircuitsynthesis-2024}, circuit optimization \cite{RL_circuitOPtimization}, and the preparation and engineering of entangled states \cite{RLStatePReparation,RLEntanglementStatePreparation,RL_entanglementEgineering}.

Quantum neural networks (QNNs) provide a flexible framework for these tasks by optimizing parametrized quantum circuits using classical learning algorithms \cite{schuld2021supervisedquantummachinelearning}. However, most existing QNN architectures rely on strictly linear parametrizations, which limits their expressive power at fixed circuit depth. Increasing depth can partially alleviate this limitation but leads to higher hardware noise and reduced scalability, making low-depth, linearly scaling architectures particularly attractive for near-term devices \cite{Luo_2023}.

A key ingredient of our approach is the introduction of different nonlinear behaviors. While nonlinear activation functions are essential for the success of classical neural networks, only a few quantum architectures incorporating nonlinearity have been explored so far \cite{QCNN-nonlinear, Reservoir-Spagnolo2022, CVNonlinearQNN, AlbertoMemristor,TripartiteEntMemristor, EntangledQuantumMemristor}. This represents a missing yet important step in statistical learning, whose role is well established \cite{Activactoins-Kunc2024}. 
The practical problem is that, while implementing an activation function is straightforward for classical networks, in the context of quantum networks, this would demand the realization of specific hardware gates and experimental solutions.

In this work, we address the challenge of engineering multipartite entanglement at fixed circuit depth under realistic noise by introducing a quantum neural network architecture with memory-inspired nonlinear parameter response. Rather than increasing depth or gate count, our approach exploits nonlinearity and circuit topology as complementary design resources to enhance the probability of generating highly entangled states.

To demonstrate the edge offered by our novel application of nonlinear functions, we run a Monte Carlo sampling over circuit architectures to showcase it right off from the noiseless case. To enable scalable optimization for mixed states, we employ the Meyer–Wallach global entanglement as a computationally efficient surrogate cost function, and certify entanglement using bipartite negativity. Our results establish a principled and experimentally motivated framework for multipartite entanglement engineering on near-term quantum devices, highlighting nonlinearity as a key ingredient for noisy algorithms.

We first demonstrate the edge offered by the non-linear activation function on a noiseless scenario, then moving to situations where the presence of noise is relevant.
Our solution can be used in two ways: (i) use the training weights directly in a quantum computing scenario or (ii) further develop this framework and implement it in an experimental setting.

The paper is organized as follows: in Section~\ref{sec:preliminaries}, we introduce the entanglement metrics, the nonlinear functions and their justification, and the gate decomposition for circuit situations. In Section~\ref{sec:methods} we define the model architecture, the training, and the metrics for the noisy scenario. Lastly, the noise gate simulates a Noise-Intermediate-Scale-Quantum (NISQ) scenario. Section~\ref{sec:results} presents the numerical results. Firstly analysing the case of pure states and then focusing on noisy settings. Lastly, Section~\ref{sec:conclusion} summarizes the conclusions and outlook.

\section{Preliminaries}
\label{sec:preliminaries}

Quantifying the degree and structure of entanglement in many-body quantum systems is essential for understanding their computational and physical capabilities \cite{PRL-multipartiteEntanglementWitnesses, PRA-multipartyEntWintessForQuantumComputing, Review_multiPartyEnt, CV_multyPArtyEnt}. 
The core idea here is to realize entangled many-body states using a nonlinear QNN, built using only one type of gate. We begin by defining the network's building blocks: the key metric for the cost function, the new activation function strategy, and the only gate we use, which is a single photon beam-splitter-inspired gate for all our architectures. Lastly, we introduce the metric and further considerations for the NISQ scenario.

\subsection{The optimization metric: Meyer-Wallach Global Entanglement.}

In this work, we focus on maximizing the global entanglement amount of the output state generated by each circuit topology.  We use the Meyer-Wallach (MW) global entanglement measure in our optimization task \cite{MeyerWallach}; this offers us two advantages by using a global validation metric : (i) it can be readily implemented inside a classical cost function and (ii) it offers an experimentally accessible and compact, permutation-invariant quantifier \cite{MW_operationalInterpretation}.
The MW is defined as follows: given a pure state $\ket{\psi}$ of an $ N$-qubit register, living in the Hilbert space $\mathcal{H} = (\mathbb{C}^2)^{\otimes N}$, the measure is defined as
\begin{equation} 
    Q(\psi) = \frac{4}{N} \sum_{i=1}^N \left( 1 - \mathrm{Tr}\! \left[ \rho_i^2 \right] \right),
    \label{eq:meyer-wallach}
\end{equation}
where $\rho_i = \mathrm{Tr}_{\{1,\ldots,N\}\setminus i} \left( \ket{\psi}\!\bra{\psi} \right)$ is the reduced density operator of the $i$th qubit. The quantity $\mathrm{Tr}[\rho_i^2]$ is the purity of the reduced state, whose deviation from unity reflects the degree of mixing induced by entanglement with the rest of the register. In noisy setting the mixing can be induced by decoherence as well. Regardless, we use the MW as a surrogate cost function, jointly certifying the genuine quantum correlations through the Entanglement Negativity, as explained in Section~\ref{sec:methods}.

By construction, the measure satisfies $0 \leq Q(\psi) \leq 1$, with one corresponding to a maximally entangled state such as the GHZ, but not a W-type state, for example. As expected, the metric vanishes if and only if the state is fully separable. 

\subsection{Nonlinear Control Reparameterization}

Generally, quantum neural networks are blocks of unitary gates trained by shifting the rotation angles $\theta_i$ for each gate. Like their classical counterpart, they combine several elementary operations acting upon different qubits to realize different architectures. What is missing with respect to the classical counterpart is the implementation of nonlinear activation functions. This is a concrete drawback and lies behind the great success of deep learning; therefore, its lack is a drawback in quantum machine learning. 
For example, a multilayer perception (MLP) reduces to a combination of matrix multiplications plus nonlinear functions right after each of them, more precisely:
\begin{equation}
    \text{\rm MLP}(\bar{\theta})[\mathbf{x}_{\rm in}] = (\sigma_n\circ W_n^{\theta_n}\circ \cdots \circ\sigma_1\circ W_1^{\theta_1})\mathbf{x}_{\rm in },
\end{equation}
with $\sigma_i$ a generic nonlinear function, $\mathbf{x}_{\rm in}$ a generic input, for a network optimizing on the set of weights parameter $\bar\theta$. In practical terms, they remap the neuron's linear output into another space. While this can be implemented relatively simply in software, realizing a physical quantum gate that implements it is an open challenge.
To tackle this setback, here we study another way around, oppositely implementing them:
\begin{align}
\label{eq:new nonlinearity}
    \mathcal{U}_{\rm NN} &=\bigotimes \mathbf{U}^n(\tilde\theta_n)\cdots\mathbf{U}^1(\tilde\theta_1),\\ \nonumber
    \tilde\theta &=\sigma(\theta).
\end{align}
As Eq.\,\,\eqref{eq:new nonlinearity} states, the activation function now modifies the output forcing it inside its dominium, and is later linearly used by a standard gate.
The first $\sigma$ we test is inspired by a photonic quantum memristor with single photon input \cite{spagnolo2022experimental,AlbertoMemristor}.
Given a physical-inspired reflectivity value $R(\theta)$, see Eq.\,\,\eqref{eq:response function}, the function simulates an effective beam-splitter rotation angle:
\begin{equation}
\tilde\theta = 2\arcsin\,\Big(\sqrt{1 - R(\theta)}\Big).
\label{eq: reflectivity}
\end{equation} 
What is physically interesting is that this type of function brings the hallmark of quantum memory effects unraveled by a memristor-type beam-splitter. In other words, even though the gate is not implemented as a genuine memristor component, it is optimized as such.

Following Eq.\,\,\eqref{eq:response function} and Eq.\,\,\eqref{eq: reflectivity}, we choose as the second activation function $\sigma_{\rm sin}$. This type of function draws inspiration from a recent breakthrough in the classical field of deep learning, dubbed Sinusoidal Representation Network (SIREN)\cite{mildenhall2022neuralfields,SIREN,SIREN-PINN, HSIREN}. Therefore, we also implement the SIREN-type nonlinearity:
\begin{equation}
    \tilde\theta = \sin(\theta).
    \label{eq: sinusoidal activaction}
\end{equation}
Lastly, to benchmark the advantage offered by the method, we consider the linear function $\tilde\theta=1\cdot \theta$.
\newline
\noindent{\textbf{Photonic quantum memristor gate decomposition -- }} The unitary decomposition in a qubit framework of the photonic quantum memristor (PQM) offered in \cite{AlbertoMemristor}, can be further decomposed using standard quantum gates \cite{SolovayKitaevTheorem}. The operation sequence is the following:
\begin{enumerate}
    
    \item A single-qubit rotation $ \mathrm{RY}$ acting on wire A, which introduces a controlled phase accumulation.
    \item An entangling sequence modeling the beam splitter dynamics,
    \[
    \mathrm{CNOT}_{\mathrm{B}\rightarrow \mathrm{A}},
    \mathrm{CRX}_{\mathrm{A}\rightarrow \mathrm{B}}(-\theta),
    \mathrm{CNOT}_{\mathrm{B}\rightarrow \mathrm{A}},
    \]
    whose combined effect implements the reflectivity-dependent partial exchange between the two modes.
    \item Lastly, a final $\mathrm{SWAP}$ gate that produce a SWAP-network entangler structure.
    \end{enumerate}
 This entire block of operations simulates the behaviour of a memristor-type beam splitter, deeply characterized in Refs.~\cite{spagnolo2022experimental, AlbertoMemristor}.  From now on, we call this block of operations $\mathbf{U}^{\rm BM}$ for a gate that implements Eq.~\eqref{eq: reflectivity} as nonlinear function, and $\mathbf{U}^{sin}$ for the one that implement Eq.~\eqref{eq: sinusoidal activaction}, and in case of linear network we use standard notation $\mathbf{U}^{\theta}$; in general we refer to it as $\plaquette$ where $\varkappa = \{ \rm BM,sin,\theta\}$.

\section{Methods}
\label{sec:methods}
\subsection{The architecture and optimization task }

Throughout this work, we make use of the $\mathbf{U}^{\varkappa}$ only to build our networks, which, in general, we define with the symbol $\mathcal{U}_{\rm topology}^{\varkappa}$, whose definition we now introduce. 
Each network possesses a specific topology, which describes how each $\bmgate$ operation (or gate) connects pairs of qubits. So, a network topology can simply be described as a list $\left[(A_1,B_1)\dots(A_n,B_n)\right]$ to apply to our $n$ gates. $A_n$ and $B_n$ are labels through which we refer to the qubits targeted by the $n$th operation.
In this work, we make use of 2 key topologies, random (RN), staircase (SC):
\begin{align}
    \text{SC} &= \plaquette_{0,1}\plaquette_{1,2}\cdot\cdot\cdot\plaquette_{n-2,n-1}\\
    \text{RN}&=\plaquette_{u_0,u_1}\plaquette_{u_1,u_2}\cdot\cdot\cdot\plaquette_{r_{n-1},r_{n}},
    \label{eq:rn topo}
\end{align}
with $n$ the number of qubits\footnote{note that for having $n$ applied gates, the labels must range from 0 to $n$}.  
The SC topology has been studied in theoretical network complexity studies to determine an optimal method to build a unitary operation from Haar-random two-qubit quantum gates \cite{HaarStaircase-Haferkamp2022} and is also the same implemented for linear graph states \cite{GraphStates}.  

\begin{figure}
    \centering
    \includegraphics[width=0.95\linewidth]{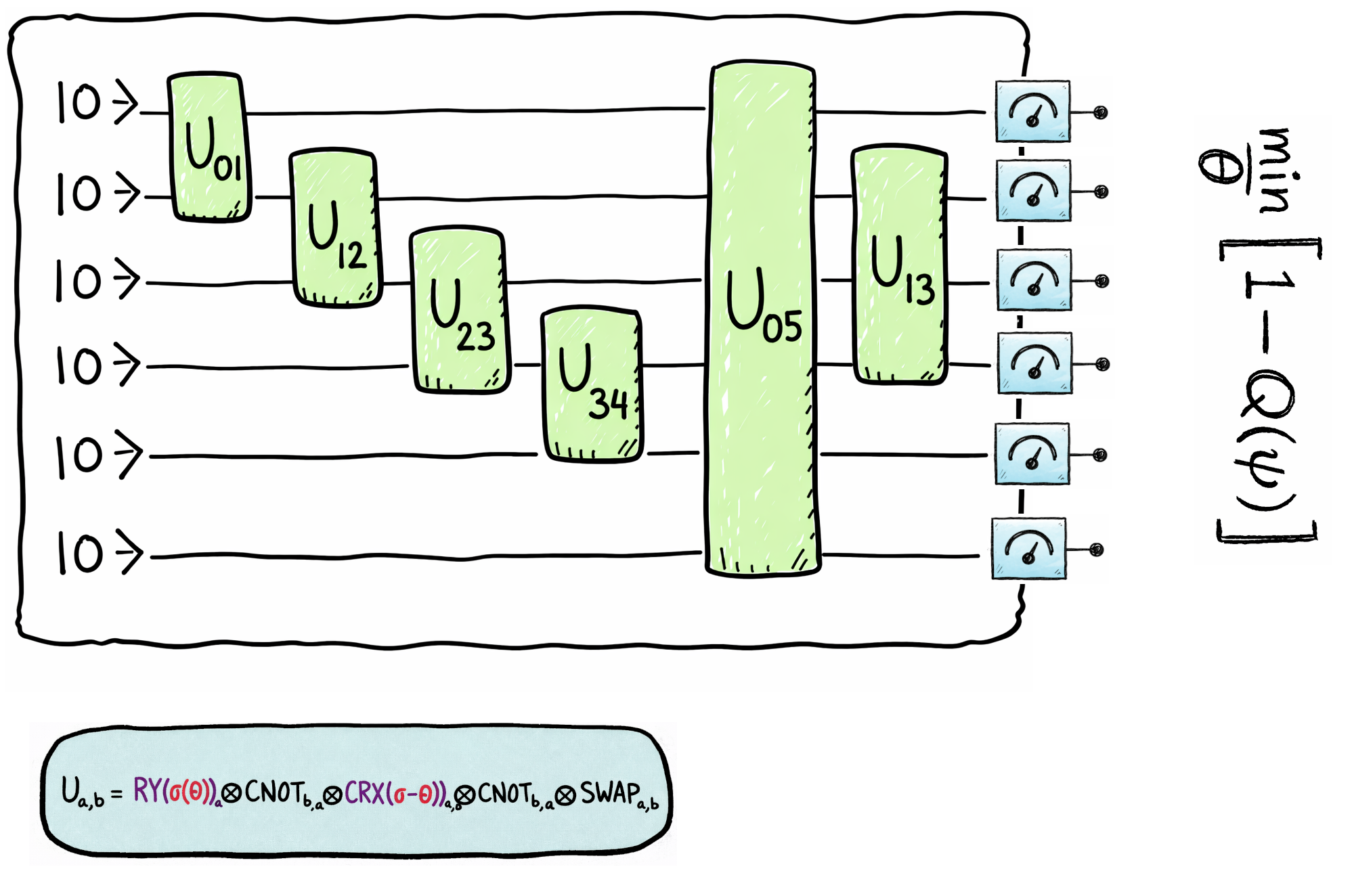}
    \caption{\justifying Schematic picture of one random network realization considered throughout our analysis. Each network is initialized from the state $\ket{0}^{\otimes n}$, and consists only of a set of gates that decompose the photonic quantum memristor gate \cite{AlbertoMemristor}; inside it, a non-linear function is applied to the variational angle of the $\mathrm{RY}$ and $\mathrm{RX}$ gates. By tuning the $\bar{\theta}$ angles, we optimize the output state generated by a specific network topology, maximizing the amount of global entanglement, as defined by the cost function \eqref{eq: nonlinear optimization}. If noise is also considered, the cost function serves as a surrogate for entanglement maximization.
    }
    \label{fig:placeholder}
\end{figure}

So, e.g.,  a network with SC topology that implements the beam-splitter type nonlinearity Eq.~\eqref{eq: reflectivity}   will be written as $\ubm_{SC} = \bigotimes \bmgate_{0,1}\cdots\bmgate_{n-2,n-1}$. We will also consider combinations of these networks, which will be defined when needed.

With this in hand, we can now define our entanglement generation problem as a minimization problem as follows:
\begin{align}
    \min_{\bar\theta}\Big[1- Q(\mathcal{U}^{\varkappa}_{\rm any}(\bar\theta)\ket{0}^{\otimes n})\Big],
    \label{eq: nonlinear optimization}
\end{align}
for \textit{any} possible topology hereafter considered.
Importantly, for the mixed state, we still employ the MW because it is a scalable, experimentally accessible surrogate cost function, whose effectiveness we validate against negativity as a certification metric.

Finally, we remark on a key takeaway: this method is not a standard supervised learning \cite{Goodfellow-et-al-2016}. Each network topology embodies a single state, so we have a 1-network $\to$ 1-state, and the protocol optimizes the set of angles maximizing the MW, together with the negativity and GME amount for each of them. 

\subsection{Primitive gate break-down}
The $\mathbf{U}^\varkappa$ gate is intrinsically an \emph{entangle-and-route} operation, because its native decomposition includes a terminal $\mathrm{SWAP}$. Consequently, if we pick one random depth-6 an RN network, the $\mathrm{SWAP}$ operations permute logical labels across physical wires, yielding a``conveyor-belt'' schedule in which one logical mode repeatedly encounters multiple partners. For example, we consider an architecture like $\mathcal{U}_{\rm sc}^\varkappa \otimes \mathbf{U}_{4,0}^\varkappa \otimes\mathbf{U}_{3,1}^\varkappa$ with optimized angles $\boldsymbol{\theta}=(0.8359,0.8665,0.2264,0.3026,0.3939,0.4087)$. Tracking the induced permutations shows that the first five applications predominantly realize successive interactions between a single itinerant logical qubit and the remaining register, while the final $U(1,3)$ introduces an additional (non-hub?) coupling (effectively a leaf-to-leaf interaction) that enhances correlation redistribution beyond the simple SC topology, for example. It is worth noticing that $\boldsymbol{\theta}$ values fall outside the Clifford set (full data for $10^5$ states available online); so, the resulting output states are expected to be non-stabilizer in general, therefore not members of the graph-state family. We thus view this architecture as implementing a SWAP-network entangler that generates weighted-graph-like, non-stabilizer states, with one state for each possible topology. The extra cross-link(s) in the RN provide a mechanism to support high global entanglement.

\subsection{Noisy simulation. Validation metric and Entanglement certification}

To depict a more realistic scenario, it is necessary to take into account the unavoidable process of decoherence in realistic quantum hardware. To do so, we include noise sources at each gate level of our qubit circuit. We implement standard hardware noise sources, namely a dephasing channel after the $\mathrm{RY}$, and an amplitude-damping after each wire of the simulated memristor. The noisy channel is defined as
\begin{eqnarray}
\mathcal{E}_{\text{deph}}(\rho) &= (1-p)\rho + p\, Z\rho Z, \\
\label{eq:dephasin}
\mathcal{E}_{\text{AD}}(\rho) &= 
K_0 \rho K_0^\dagger + K_1 \rho K_1^\dagger,
\label{eq:damping}
\end{eqnarray}
with $Z$ the standard Pauli operator, and the Kraus matrices for the amplitude damping channel given by
\begin{align}
K_0 &= 
\begin{pmatrix}
1 & 0 \\
0 & \sqrt{1-\gamma}
\end{pmatrix},
&
K_1 &= 
\begin{pmatrix}
0 & \sqrt{\gamma} \\
0 & 0
\end{pmatrix}.
\end{align}
To quantify the entanglement of the mixed output states, we use Entanglement negativity \cite{Review-Negativity}, which is a LOCC monotone, and use the positive-partial-transpose criterion \cite{PPT_criterion,PPTmixed_HORODECKI1996}.
Given a bipartite density matrix $\rho_{AB}$, one compute the partial trace
with respect to subsystem $B$, obtaining $\rho_{AB}^{T_B}$.  
The entanglement negativity is then defined as
\begin{equation}
\mathcal{N}(\rho_{AB}) = \frac{\|\rho_{AB}^{T_B}\|_1 - 1}{2},
\end{equation}
where $\|\cdot\|_1$ denotes the trace norm.  
Regarding the MW global entanglement, this is no longer interpreted as a strict entanglement monotone and an entanglement validation metric, but rather as a scalable proxy and a good surrogate cost function, promoting multipartite correlations during network optimization.

Lastly, we certify genuine multipartite entanglement (GME) in mixed states using the PPT-mixture semidefinite-programming (SDP) criterion~\cite{TamingEntanglement} on mixed states. A description of the method is provided in Appendix\,\ref{app:SDP} and goes beyond the scope of this work.

\noindent{\bf Training hyperparameters -- } For the optimization, we implement the classical Adam optimization, a learning rate of $5\cdot 10^{-3}$, and early stopping for regularization.   $\mathrm{Pennylane}$ is used to code and train all the models\,\,\cite{PennyLane}.

\section{Results}
\label{sec:results}

The first mandatory question to assess is the boost offered by nonlinearities inside our optimization scheme. A Monte Carlo analysis of this key question for the noiseless pure-states scenario is offered. The main outcomes hereafter displayed will leverage the numerical evidence and focus on a step-by-step analysis, starting from 5 low-noise qubits up to 10 in NISQ scenario. 
\begin{figure}[h!]
    \centering
    \includegraphics[width=1\linewidth]{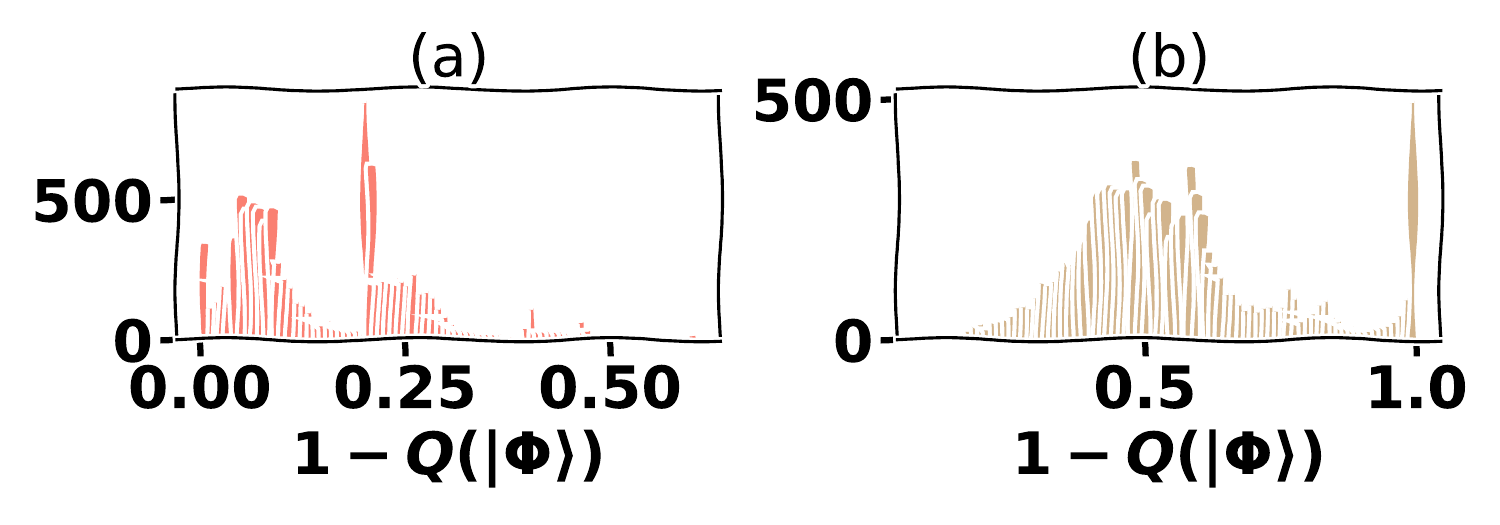}
    \caption{\justifying (a) Global entanglement histogram of $10^5$ different random topologies $\mathcal{U}^{\rm BM}$realizations, after optimization for fixed depth five. (b) Same histogram but for $10^5$ $\mathcal{U}^\theta$ random topologies. In both cases no noise was considered. This demonstrate the neat training advantage offered by the presence of a non-linear.
    }
    \label{fig: 5 monte carlo}
\end{figure}

\noindent{\bf Activation functions advantage -- }
The histograms in Fig.\,\,\ref{fig: 5 monte carlo} displays the validity of our hypothesis: (i) from the Monte Carlo analysis of $10^5$ random noiseless circuits, we show that the use of nonlinear functions offers a significant boost for the entanglement production, and (ii) in this regime, because the Meyer-Wallach is both an entanglement cost function and a valid validation metric, we can scale up to 20 qubits simply by implementing the SC topology for almost pure states, see Fig.~\ref{fig:20 qubits}.

\subsection{5 qubits on low noise analysis}

\begin{figure}[t!]
    \centering
    \includegraphics[width=0.92\linewidth]{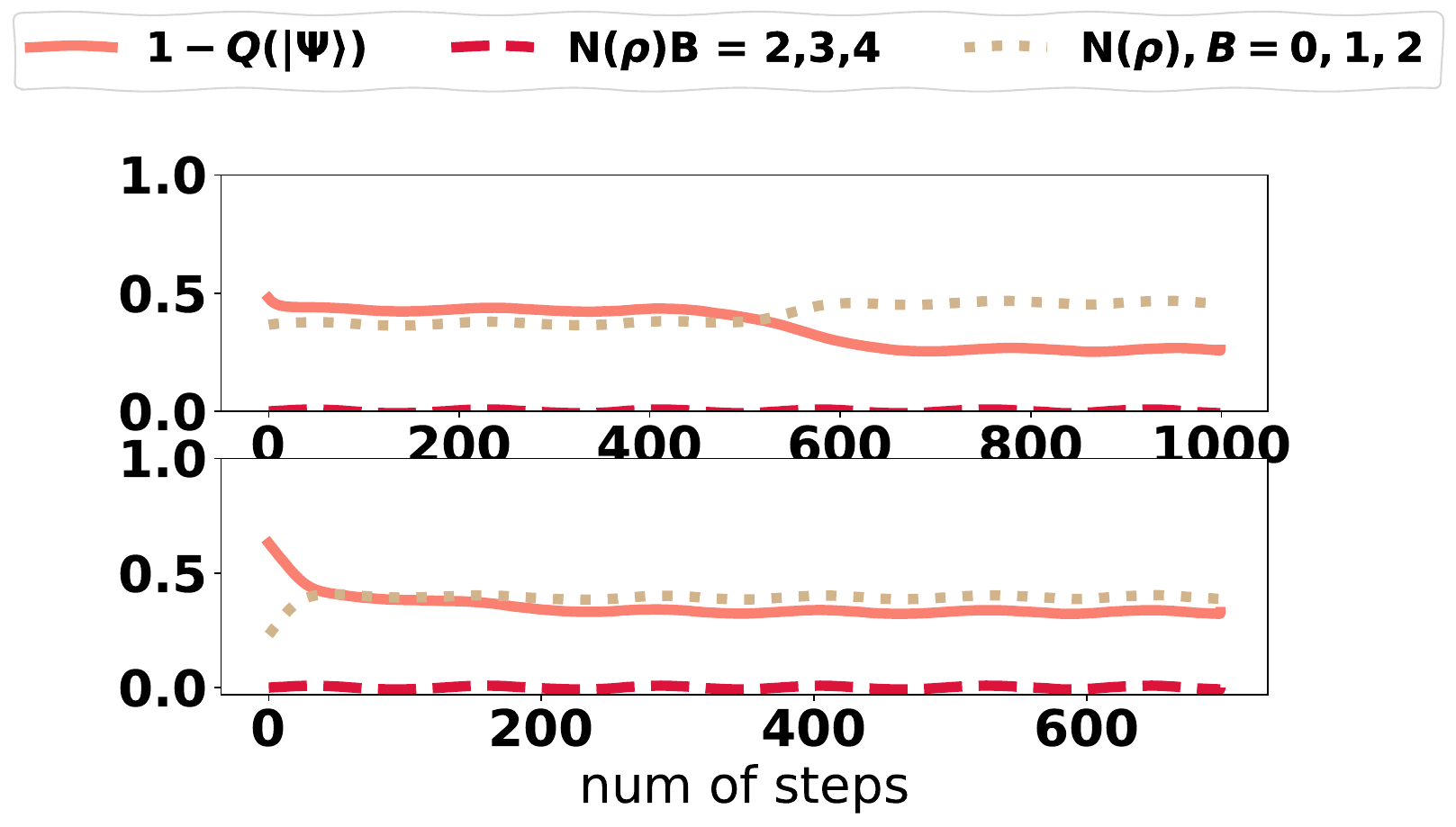}
    \caption{\justifying Upper panel. Meyer-Wallach and bipartite negativity for a single run of the $\usc^{\rm BM}$ models, i.e, the simple staircase model. Lower panel, the same plot for $\usc^{\rm sin}$ model. We can see that the optimization of this topology produces an amount of bipartite mixed state entanglement $\sim 0.5$, for both bipartitions, also in the presence of very low noise. This clearly prompts us to introduce more gates between non-nearest-neighbor qubits. }
    \label{fig:sc with amplitude damping}
\end{figure}

In this section, we analyze the networks $\ubm_{\rm sc}$ and $\ubm_{\rm sc+1}$ under low noise. We implement Eq.~\eqref{eq:damping} to introduce an amplitude damping gate at the end of each SWAP's wires, with a $\gamma= 0.01$, 
our numerical experiment shows that the SC topology poorly performs in this regime. Not only does the cost function minimization -- which is equivalent to the Meyer Wallach maximization -- not go to zero. Also, the amount of entanglement in the sub-partition is $<0.5$, a low value compared to the theoretical upper bound of $\sim 1.5$. Instead, by adding an extra connection on the qubits' wires $(0,4)$ (which we define as topology SC+1), the situation changes. An example run of such a setup is shown in Fig.~\ref{fig:sc with amplitude damping}.

The negativity is provided for the partition $ [0,1,2]\,\text{and}\,[2,3,4]$ for an initial state  $\ket{\psi_0} = \ket{0}^{\otimes 5}$. The theoretical upper bound for this partition is, for a pure state $\sim1.5$  \cite{Review-Negativity}.

\begin{figure}
    \centering
    \includegraphics[width=0.92\linewidth]{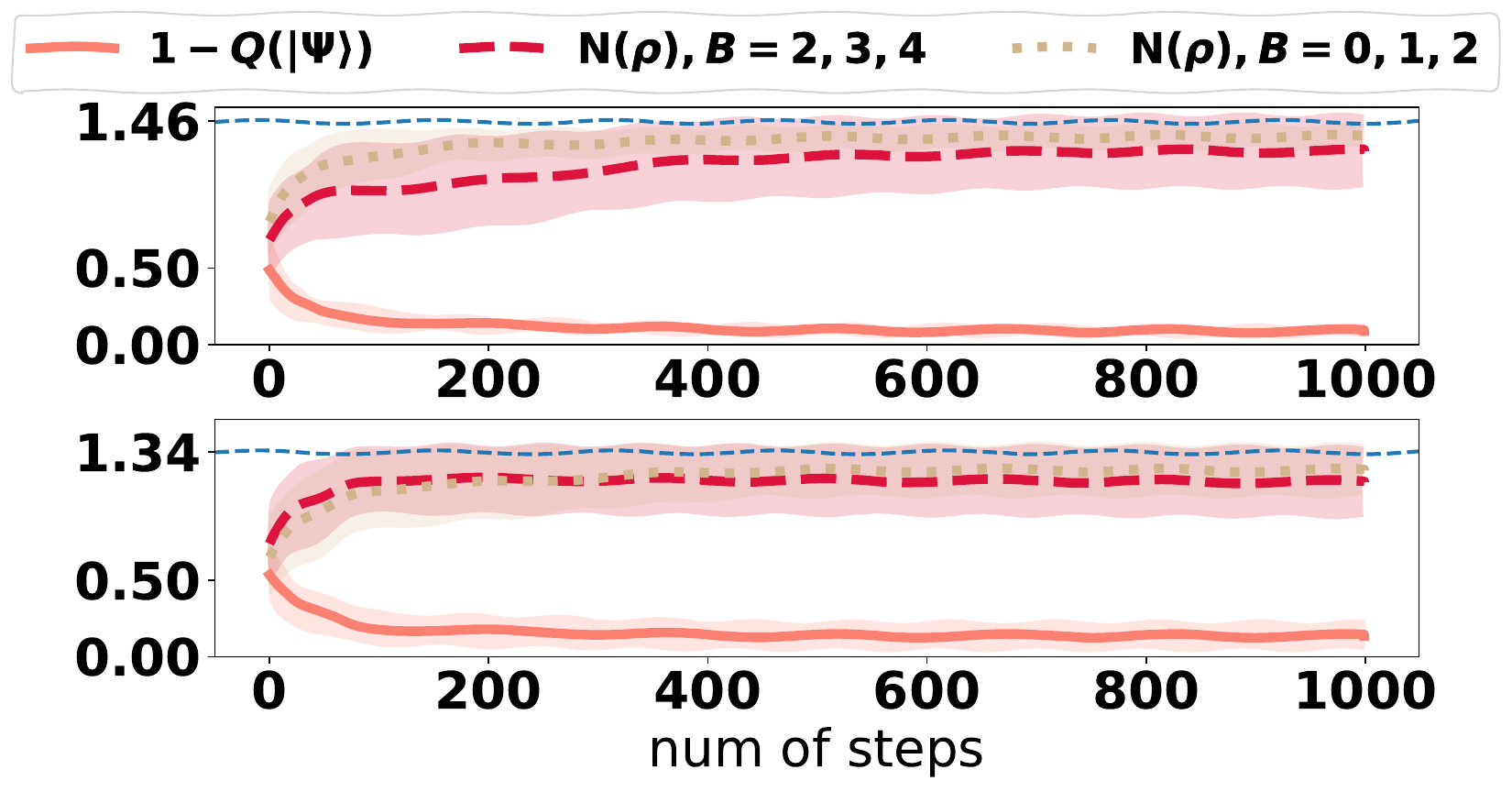}
    
    \caption{\justifying Upper panel. Mean value and standard deviation training evolution for 20 different random initializations of the network $\uscone^{BM}$. Lower panel. The same for $\uscone^{sin}$ networks. In the upper panel, we can see that the memristor-inspired architecture achieves a higher standard deviation, but can potentially obtain a close-to-theoretical value for B=[0,1,2],  while this is not the case for the $\uscone^{\rm sin}$ model. This shows the potential and the sensitivity as well to the network's initialization values.}
\label{fig:partial noise 5q stat}  
\end{figure}

Undoubtedly,  Fig.~\ref{fig:partial noise 5q stat} confirms a degree of sensitivity of the network to initial parameter weights, as showcased by the high variance. Looking at the upper panel, we can see that  $\mathcal{U}^{\rm BM}_{\rm sc+1}$ is bound to achieve greater performance, a value of $\sim 1.46$ when the theoretical upper bound is $1.5$, demonstrating the efficiency of the approach for a small number of qubits. We can appreciate now how the $\uscone^{\rm BM/sin}$ models' standard deviation in the learning curve is evident now, around $\sim10^{-1}$.

\subsection{5 qubits NISQ scenario}
\label{sec:nisq}

To investigate the performance of the proposed architecture under realistic near-term conditions, we simulate a NISQ scenario by including both dephasing and amplitude-damping noise at the gate level. Specifically, a dephasing channel is applied immediately after each single-qubit rotation $R_Y$ within every $U_\kappa$ block, while amplitude damping with rate $\gamma = 0.01$ is applied after each wire of the simulated beam-splitter operation.

As stated in Sec.\,\,\ref{sec:methods}, the MW metric is now used as a valid cost function and scalable surrogate, and the multi-partite entanglement is certified with bipartite negativity.

The presence of relevant noise cannot help but reduce the amount of generated entanglement. We remind that the $\mathrm{SWAP}$ belt-conveyor topology redistributes the entanglements across the network;
therefore, we consider a circuit with an extra block:
\begin{equation}
\label{eq:staircase plus two}
\mathcal{U}_{\mathrm{SC+2}}^{\mathrm{BM/sin}} =
\mathcal{U}_{\mathrm{SC+1}}^{\mathrm{BM/sin}} \otimes \mathbf{U}^{\mathrm{BM/sin}}_{3,1},
\end{equation}
which increases the number of long-range interactions.

\begin{figure}
    \centering
    \includegraphics[width=0.92\linewidth]{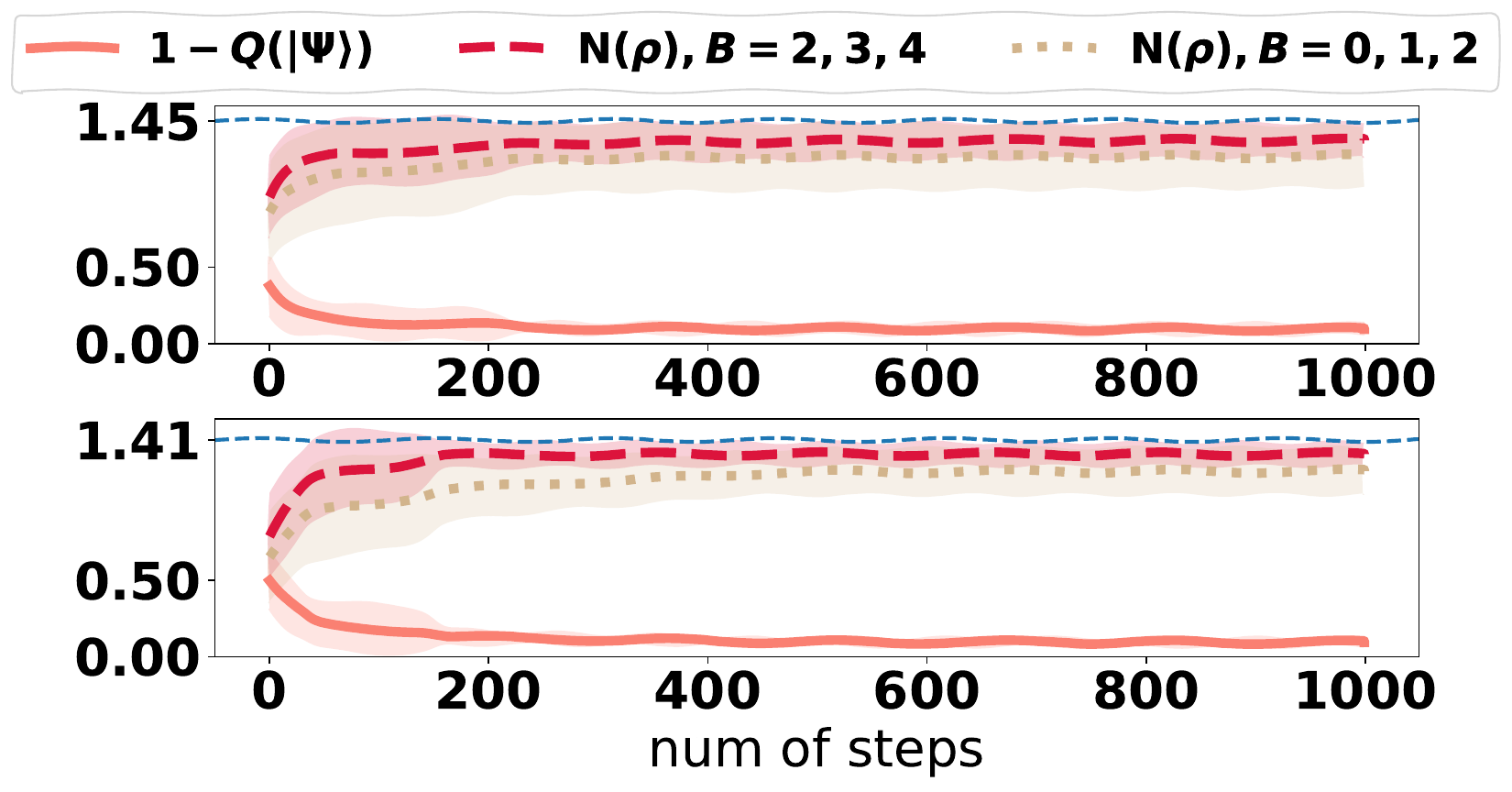}
    \caption{\justifying Mean value and standard deviation of the evolution of the full noisy simulation with dephasing channel after $\mathrm{RY}$ and amplitude damping after $U^{\varkappa}$, for 20 different networks' random initializations. Upper panel and lower panels showing the BM and sin simulations as usual. We can observe that the $\mathcal{U}_{\rm sc+2}^{\varkappa}$ can potentially achieve close to the theoretical value, and the physical memristor-inspired $\mathcal{U}_{\rm sc+2}^{\rm BM}$ is bound to be slightly more efficient. This confirms the networks' sensitivity to parameter initialization, but in contrast to Fig.\,\,\ref{fig:partial noise 5q stat}, this now reduces, as in the present figure, the standard deviation is smaller. This can be interpreted as a positive side effect of noise injection inside the differential landscape.}
    \label{fig:staircase plus funnel full noise 5q}
\end{figure}

As illustrated in Fig.\ref{fig:staircase plus funnel full noise 5q}, the additional coupling enables either smooth optimization and comparable negativity values across complementary bipartitions. In particular, for the BM nonlinear activation, the achieved negativity approaches the corresponding theoretical upper bound for the considered partitions, indicating the effective generation of multipartite entanglement despite the presence of noise.

These results highlight the crucial role of circuit topology in the noisy regime. While the surrogate Meyer-Wallach cost function drives the optimization, the network's ability to translate this optimization into genuine multipartite entanglement, as certified by negativity, depends sensitively on the connectivity structure. Enhanced topologies mitigate the adverse effects of local noise by facilitating entanglement redistribution across the many-body system.




\begin{figure}
    \centering
    \includegraphics[width=0.85\linewidth]{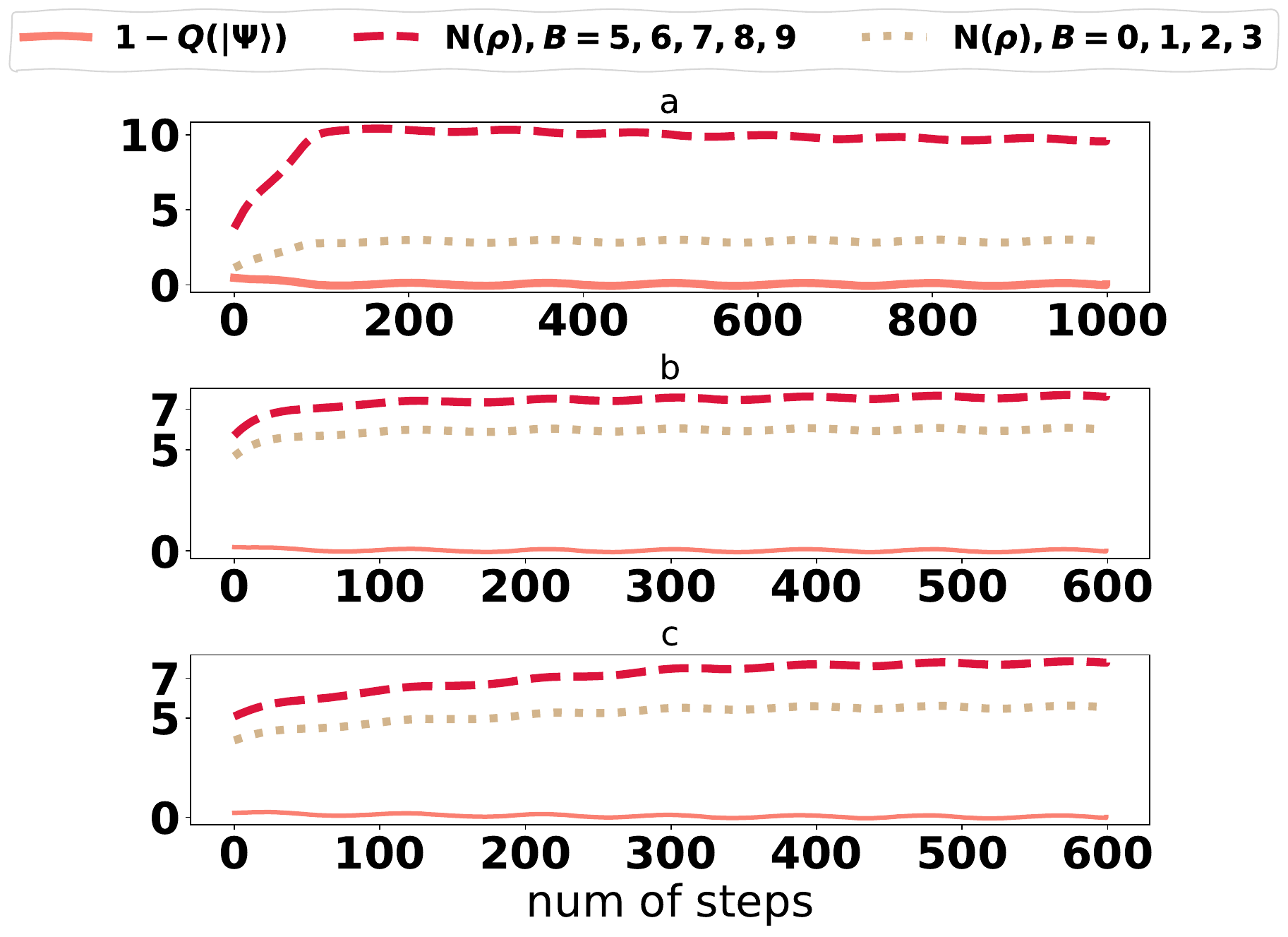}
    \caption{\justifying Panel (a). Negativity bipartite entanglement production for the $\mathcal{U}^{[5-9]}_{10}$ network, with dephasing noise of strength $p=0.01$ and amplitude damping of $\gamma=0.01$. For this topology, the production of entanglement for the mixed case achieves a value of $\sim 10$ for a theoretical maximum of 15. Panel (b) and (c). Mixed states entanglement engineering for sub partition B=[0,1,2,3], using the network $\mathcal{U}^{[0-3]}_{10}$ and $\mathcal{W}^{[0-3]}_{10}$. The topology can still produce a measurable amount of entanglement for the desired bipartition, but with more effort, and obtains in panel (b) at most a value of 6 for negativity and 5.6 in panel (c). This demonstrates that a higher amount of connectivity does not imply a stronger entanglement generation, but it depends strictly on the way we cut the state into two partitions.
    }
    \label{fig:10 qubits full noise}
\end{figure}

In Fig.\,\ref{fig:staircase plus funnel full noise 5q} we can appreciate the effects of this topology in Eq.\,\,\eqref{eq:staircase plus two}, whose structure is meant to stitch together the opposite qubits of our many-body state. Not only is the training smooth and faster, but, as can be noticed in the upper panel for $\mathcal{U}_{\rm sc+2}^{\rm BM}$, the two bipartitions can achieve a value of negativity close to the theoretical maximum of $\sim 1.5$.

\subsection{10 qubits NISQ scenario. Topology study}
Now we scale the problem for a many-body state of 10 qubits. We opt for testing three types of networks for $sine$ only nonlinear term:

\begin{align*}
    \mathcal{U}_{10}^{[5-9]} &= \uscone\mathbf{U}_{8,1} \mathbf{U}_{7,2}\cdots\mathbf{U}_{5,4}\\
    \mathcal{U}_{10}^{[0-3]} &= \uscone\mathbf{U}_{8,1}\mathbf{U}_{7,2}\mathbf{U}_{6,3}\mathbf{U}_{4,1}\mathbf{U}_{5,2}\\
    \mathcal{W}_{10}^{[0-3]} &= \mathcal{U}_{10}^{[0-3]}\mathbf{U}_{9,1}\mathbf{U}_{8,0}
\end{align*}

In Fig.~\ref{fig:10 qubits full noise}, we can see to what extent the topology design can influence the production of entanglement between bipartitions. In plot (a), we can see the action of $\mathcal{U}_{10}^{[5-9]}$. Our engineering approach is particularly effective for symmetric bi-partition. Forcing the entanglement into B= [0,1,2,3], instead, is more difficult. Comparing panel (b) with (c), we can see how the network $\mathcal{U}_{10}^{[0-3]} $ and $\mathcal{W}_{10}^{[0-3]}$ redistribute it. It is worth to highlight is that $\mathcal{U}_{10}^{[0-3]} $, in panel (b), has fewer gates, 12 instead of 14, but has better performance, obtaining $\sim6$  of negativity against the $\sim5.6$ from $\mathcal{W}_{10}^{[0-3]} $. So, for non-symmetric bi-partitions, increasing the number of gates is not fruitful.

The topology distributes the entanglement between the sub-partition B = [0,1,2,3,4], which can achieve a value of $\sim10$ for a theoretical upper-bound of $\sim 15$, while for B = [0,1,2] will increase by a relatively small value, reaching a negativity value around $\sim3$. This means that the topological structure affects the many-body properties of the outcome state.

Last, we certify genuine multipartite entanglement by excluding biseparability using the PPT-mixture SDP criterion \cite{TamingEntanglement}. The SDP becomes infeasible already when restricting to 1|9 bipartitions, which implies that the state is not a PPT mixture and therefore is GME.

\section{Conclusion}
\label{sec:conclusion}

We have introduced a new optimization approach to introduce non-linear components inside quantum neural networks, where now the non-linear function is contained in the neuron function, doing the opposite. As a testbed for our approach, we optimize the amount of global entanglement inside a quantum neural network.  To design it, we draw inspiration from experimentally available memristor photonic components, a type of gate with demonstrated ability in experimental generation of entanglement. To simulate the memristive behaviour, we physical reflectivity function obtained for a wave-shaped input. To demonstrate the generality of the approach, we also consider a simpler sinusoidal function, bridging with the classical counterpart SIREN networks. By optimizing 5-qubit  $10^5$ random topologies, we illustrate the boost from the non-linearity application. 
The analysis over the 5-qubit topology, for two different non-linear functions, offers an insight: an entanglement convey-belt, i.e., more gates applied to far-away qubits, tackles noise. Using this as a key observation, we pivot to a 10-qubit simulation for a noisy scenario and 3 different topologies, measuring the negativity for selected bipartitions. The approach can generate entanglement for mixed states, and to certify that it is genuinely multipartite, by applying the PPT-mixture semi-definite programming algorithm.

Taken together, our results establish a lightweight and experimentally motivated framework for engineering multipartite entanglement on near-term photonic platforms, in both pure and noisy quantum systems. 
Several future developments can naturally branch off, e.g., studying different metrics, best suited for mixed states, or non-linear functions,  moving to the qudit case, together with an in-depth study of experimental realization. 

\begin{acknowledgments}
A.M.P. and R.L.F acknowledge support by MUR (Ministero dell'Università e della Ricerca) through the PNRR Project ICON-Q-Partenariato EstesoNQSTI - PE00000023 - Spoke 2 - CUP: J13C22000680006.
\end{acknowledgments}
\bibliography{cites}
\newpage
\appendix

\subsection{Pure states random topologies analysis}
\label{sec:pure states}
Because each different topology can give rise to a different output state, the first question to answer is ''how many useful topologies may we expect?'' To dive into this, we run an exploratory Monte Carlo analysis sampling $10^5$ random (rm) topologies, for a 5-qubit and $\ubm_{\rm RN}$ and $\mathcal{U}^{\theta}_{\rm RN}$ models.
\begin{figure}[h]
    \centering
    
    \includegraphics[width = 1\linewidth]{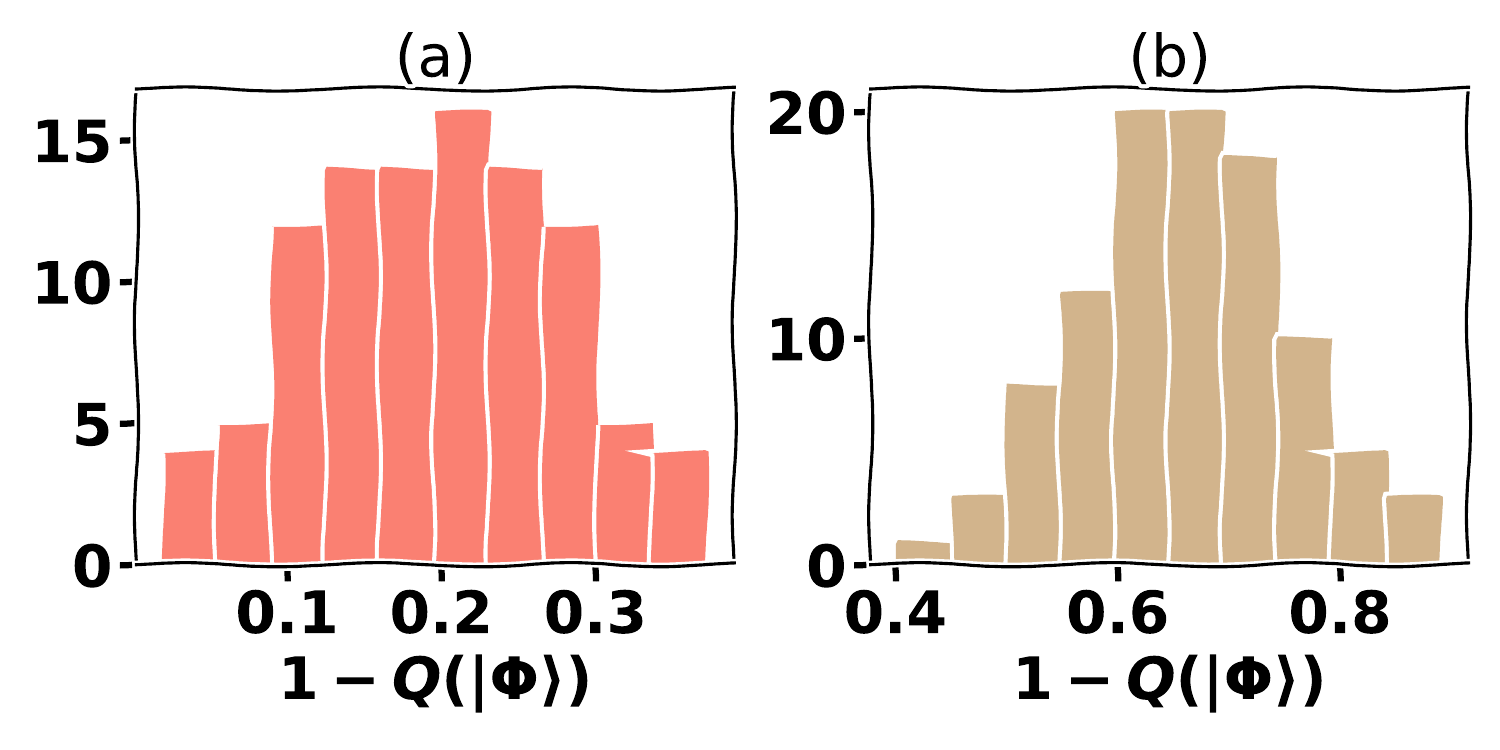}
    \includegraphics[width = 1\linewidth]{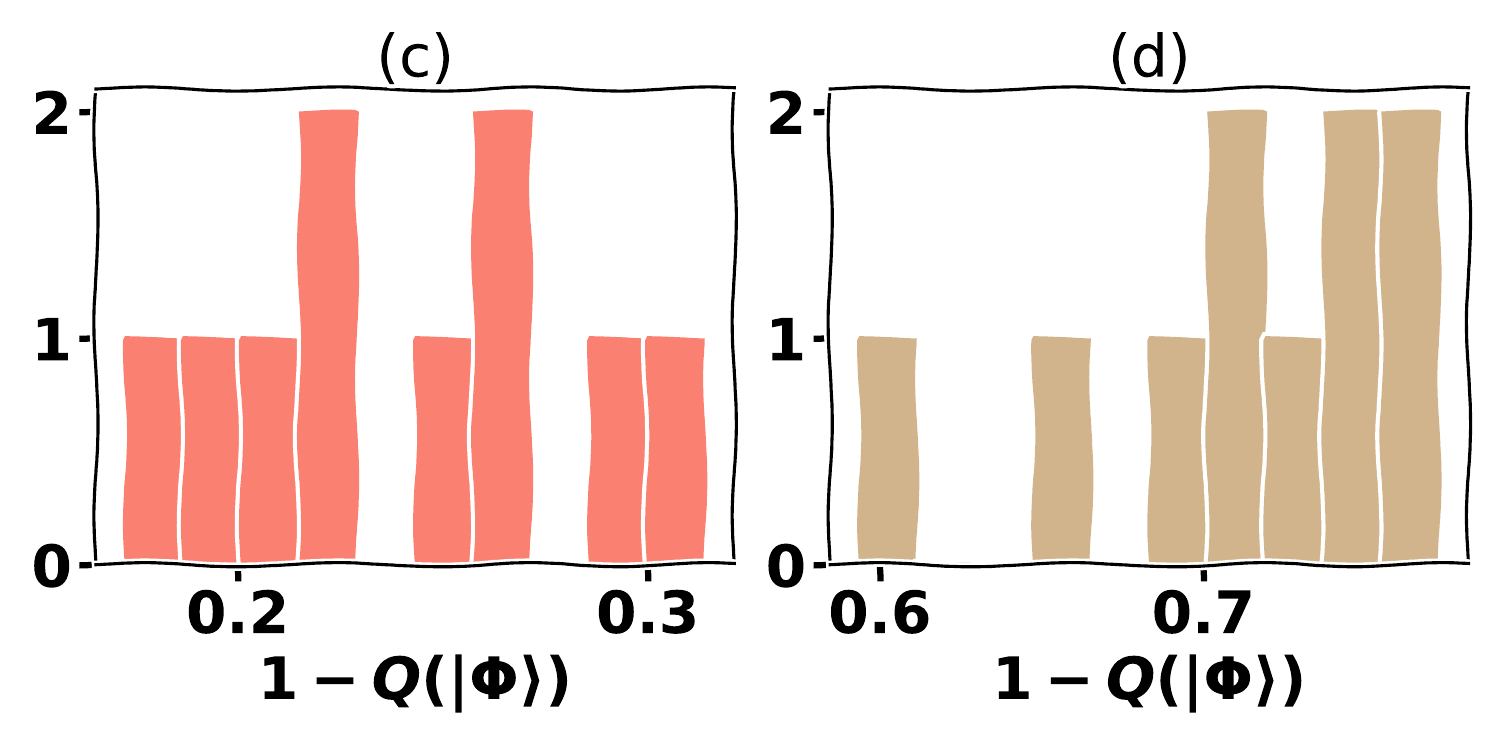}
    \caption{\justifying Panels (c) (d), histogram for 11-qubit $\mathcal{U}^{\rm BM}_{\rm rm}$ and $\mathcal{U}^{\theta}_{\rm rm}$. We compare network topologies fixing the depth equal to 11 for all random topologies. The network loss function maximizes the MW values of the output states by minimizing Eq.~\ref{eq: nonlinear optimization}. So, values close to zero in the loss function indicate higher values of global entanglement in the output pure state. We can clearly see that the presence of a nonlinearity as given by Eq.~\ref{eq:integral_exact} response function, unlocks a greater optimal minima search.}
    \label{fig:monte carlo}
\end{figure}
As Fig.\ref{fig:monte carlo} unfolds, the nonlinear activation functions allow better outcomes, with many more random QNN that can achieve values of MW above 0.8. 
This result can be leveraged to prepare a training dataset for classical deep learning solutions meant to synthesize optimal topologies. We also run a quick analysis for a small noise scenario, where only amplitude damping is considered. Again, we confirm that the presence of nonlinear functions is necessary to achieve higher performances; the results are shown in Sec.\ref{Sec:5 noisy qubits}.
Another natural question that arises is the dependency of the networks on initial conditions and parameter variation, i.e., the $T_{osc}/T_{int}$ that describes the physical gate, even though we are in a noiseless case. Results of the noiseless study can be found in \ref{Sec:initial conditions} and show that for this scenario, no dependency is found.

\noindent{\textbf{The topologies of 0.8 threshold --}} As shown in panel (a), the MW value of 0.8 turns out as a threshold, or a specific boundary when a transition occurs. We analyze the structure of topologies below 0.8 and found that all of them have at least one wire set (A,B) with $A>B$ and that connect qubits that are not neighbors. This will play a critical role in Sec.\,\,\ref{sec:nisq} when the presence of noise will be investigated.

\subsection{Scale-up and the sinusoidal hypothesis } The nonlinear function in Eq.~\eqref{eq: reflectivity} records the memory behaviour of a quantum memristor element, when a sinusoidal input is used in input. Here, we want to both (i) scale-up the problem and (ii) compare the model $\uscone^{\rm BM}$ versus $\uscone^{\rm sin}$, bridging with SIREN classical networks. To address both questions, we display the learning curve for $11\mathcal {T}_{\rm ladder}$. For this experiment, we consider two topologies $\uscone^{\rm BM/sin}= \usc^{\rm BM/sin}\otimes\mathbf{U}^{\rm BM/sin}_{10,0}$ and $\mathcal{W}_{\rm sc+1}^{\rm BM/sin}=\usc\otimes\mathbf{U}^{\rm BM/sin}_{0,10}$.
\begin{figure}
    \centering
    \includegraphics[width=0.8\linewidth]{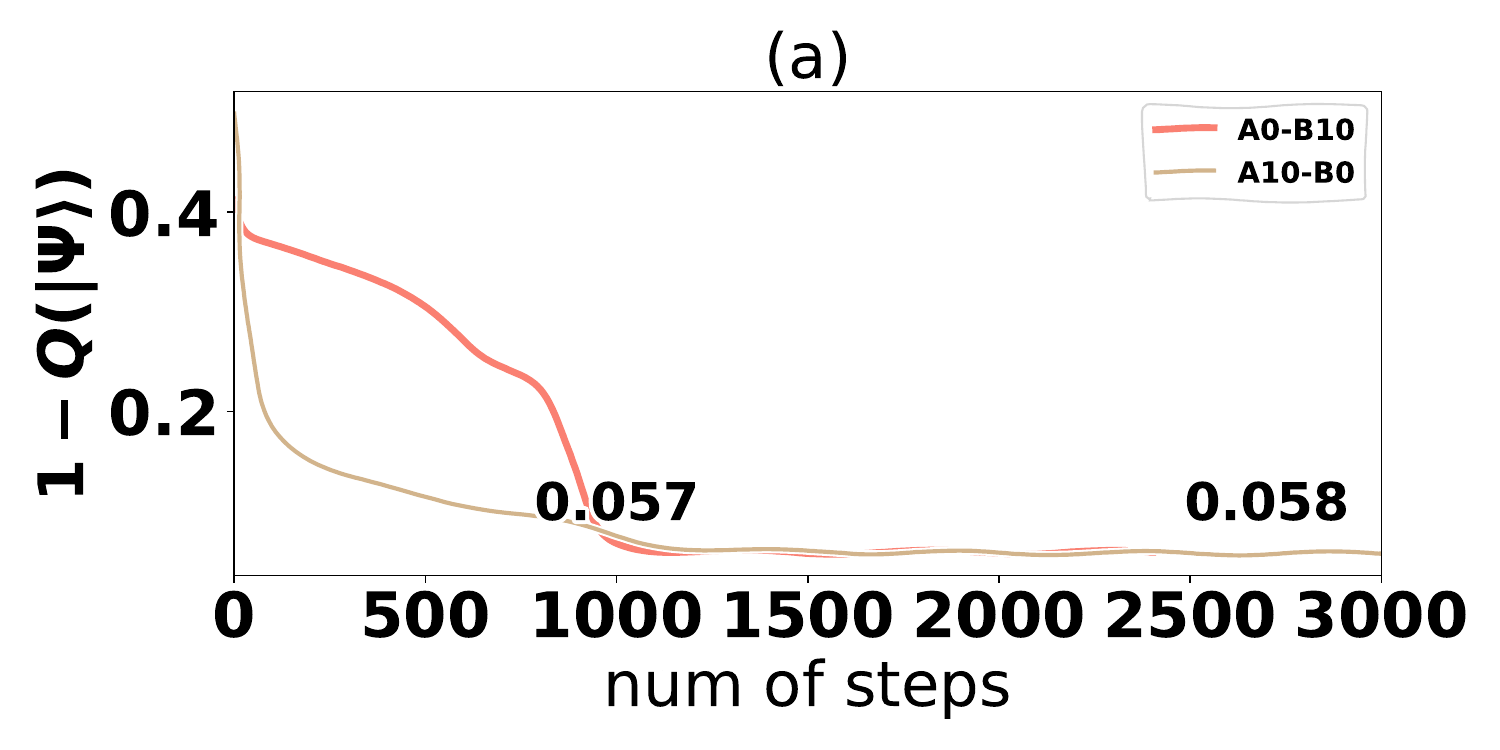}
     \includegraphics[width=0.8
     \linewidth]{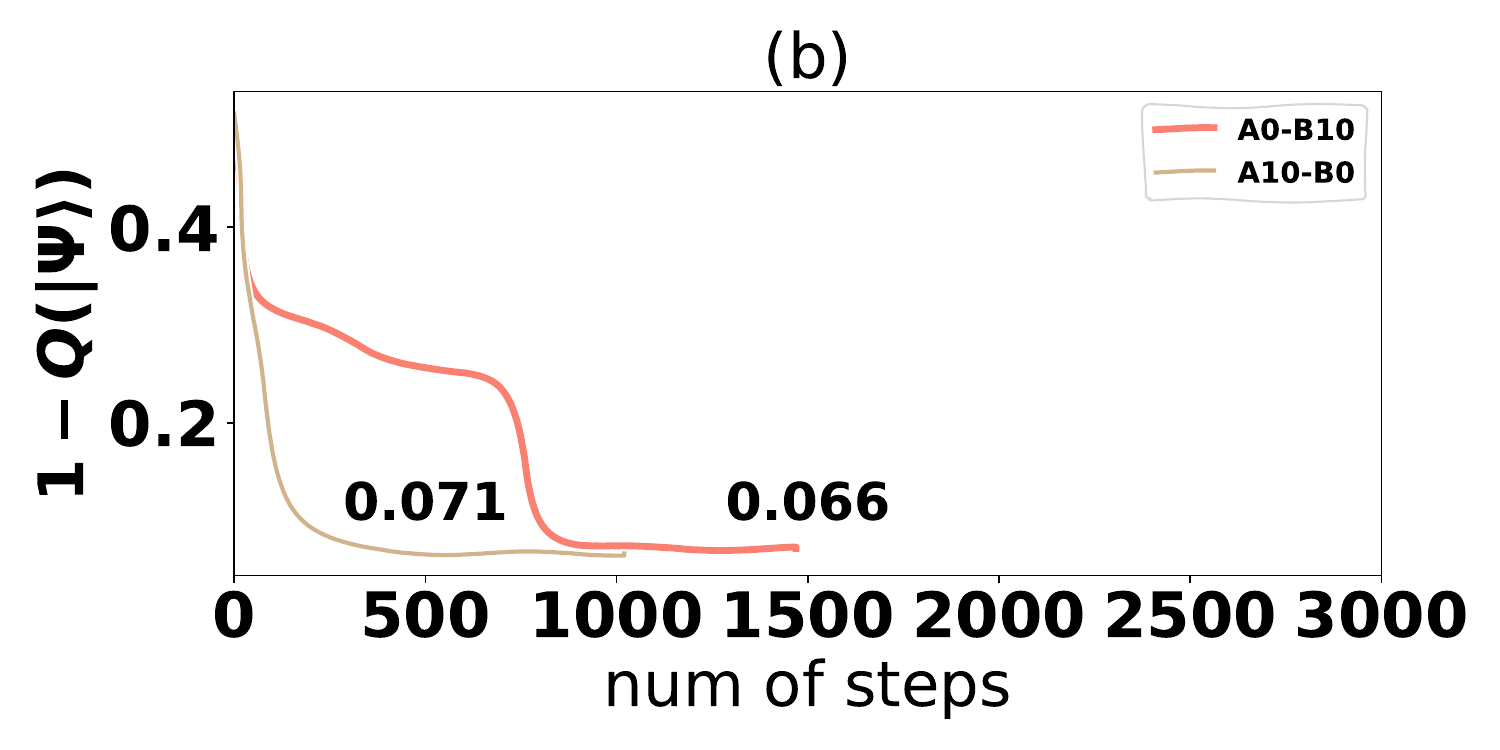}
    \caption{\justifying Panel (a). Optimization history -- the Meyer-Wallach entanglement maximization -- for the $\uscone^{\rm BM}$ and $\mathcal{W}_{\rm sc+1}^{\rm BM}$  architecture with  for a 11 qubits quantum network. Panel (b). The $\uscone^{\rm sin}$ and $\mathcal{W}_{\rm sc+1}^{\rm  sin}$ architectures outcomes. The beam-splitter nonlinearity term brings a slightly better performance, but the $sin$ term leads to a faster convergence. This may point out a better fitness of the $sine$ nonlinear term, for general cases, like the noisy one.}
    \label{fig:11 qubits}
\end{figure}

For 11 qubits, Eq.~\ref{eq: reflectivity} and Eq.~\ref{eq: sinusoidal activaction} are numerically equivalent. This means that sinusoidal nonlinear functions are a valuable tool in our task of entanglement engineering, and the analogy with the SIREN is well-posed. The only appreciable difference between the two terms is the number of epochs (steps) needed to achieve convergence. The $sine$ function offers a faster convergence; this can be the hallmark of a potential term with better fitness to our optimization problem. To conclude, we run the same analysis for 20 qubits either; this can be found in \ref{sec:20 qubits noiseless}.

\section{Initial condition dependency } 
\label{Sec:initial conditions}
For both quantum and classical models, if the differential landscape is not smooth, the solution may oscillate due to the initial set of random weights use to initialize them. In this section we study how changes in the leading physical values of Eq.~\ref{eq:response function} may affect the final outcomes of the network.
We choose $T_{\rm osc}/ T_{\rm osc} = [0.2,0.5,0.8,1.1,1.4] $. We fix the network choosing $\mathcal{U}_{\rm ladder}$, both noise and noiseless, then run for each pair $(T_{\rm osc},T_{\rm int})$ 20 optimization runs, to calculate the mean and variance obtained for each combination. The key result is that the variance, in both noisy and noiseless scenarios, is below $10^{-8}$, while the mean value ranges between $0.07-0.03$ and $0.11-0.06$ for the noiseless and noisy scenarios, respectively. This is likely due to the linear complexity of the network, which needs to optimize a small number of parameters. Therefore, the differential landscape is very smooth.

\section{5 noisy gates analysis}
\label{Sec:5 noisy qubits}

\begin{figure}[h!]
    \centering
    \includegraphics[width=0.9\linewidth]{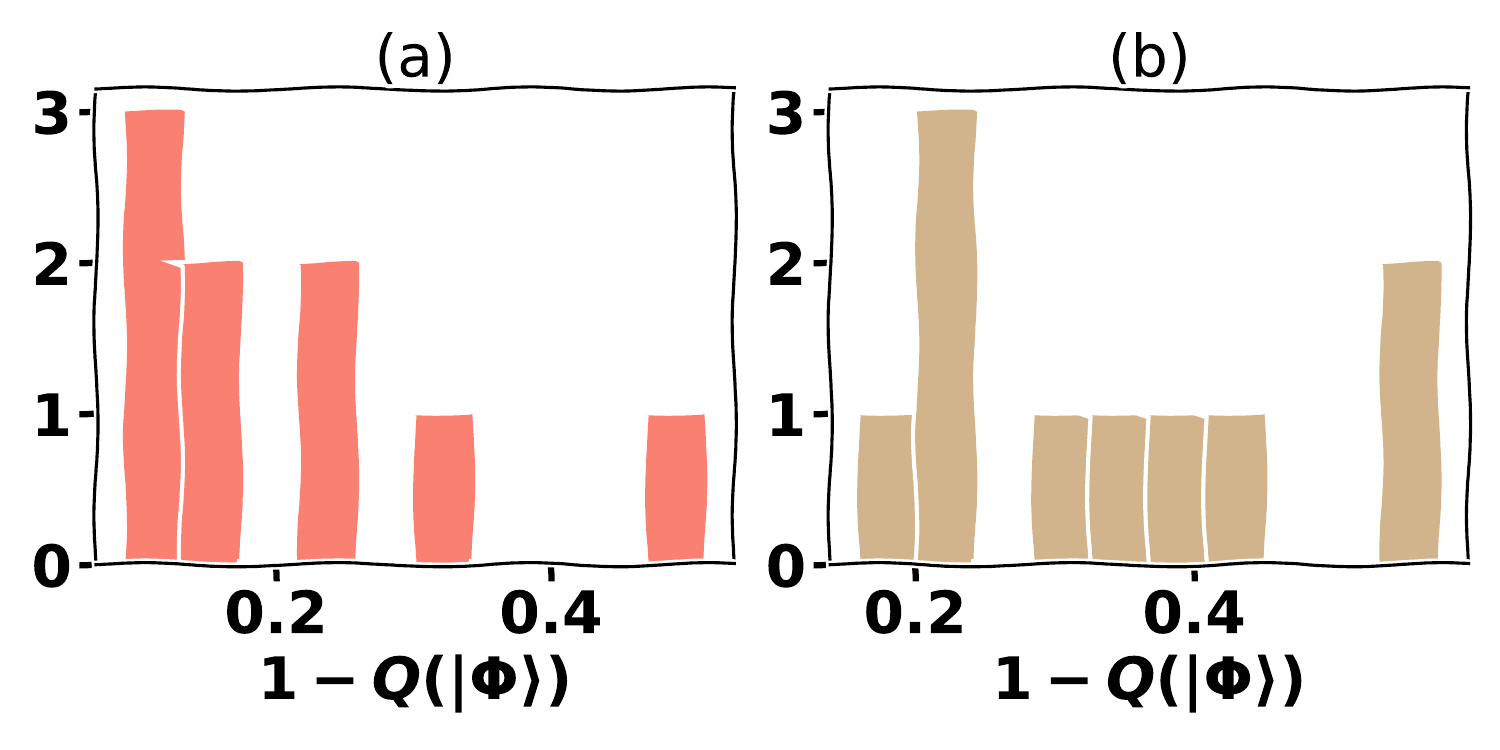}
    \caption{\justifying Histogram for 10 different random topologies for a 5-qubit neural network. On the left, the $\mathcal{U}^{\rm sin}_{\rm SN}$ structure, and on the right, the $\mathcal{U}^{\rm \theta}_{\rm RN}$ model sharing same RN topologies. The sin model is pretty much equivalent to the beam-splitter, but always superior to the linear one.}
    \label{fig:noisysin montecarlo}
\end{figure}

In this last round of experiments, we consider the situation where an amount of photon loss, here modeled using an amplitude damping channel with $\gamma=0.001$, is added at the end of each gate on both wires. We sample ten random topologies and use them to run two
\begin{figure}
    \centering
    \includegraphics[width=0.9\linewidth]{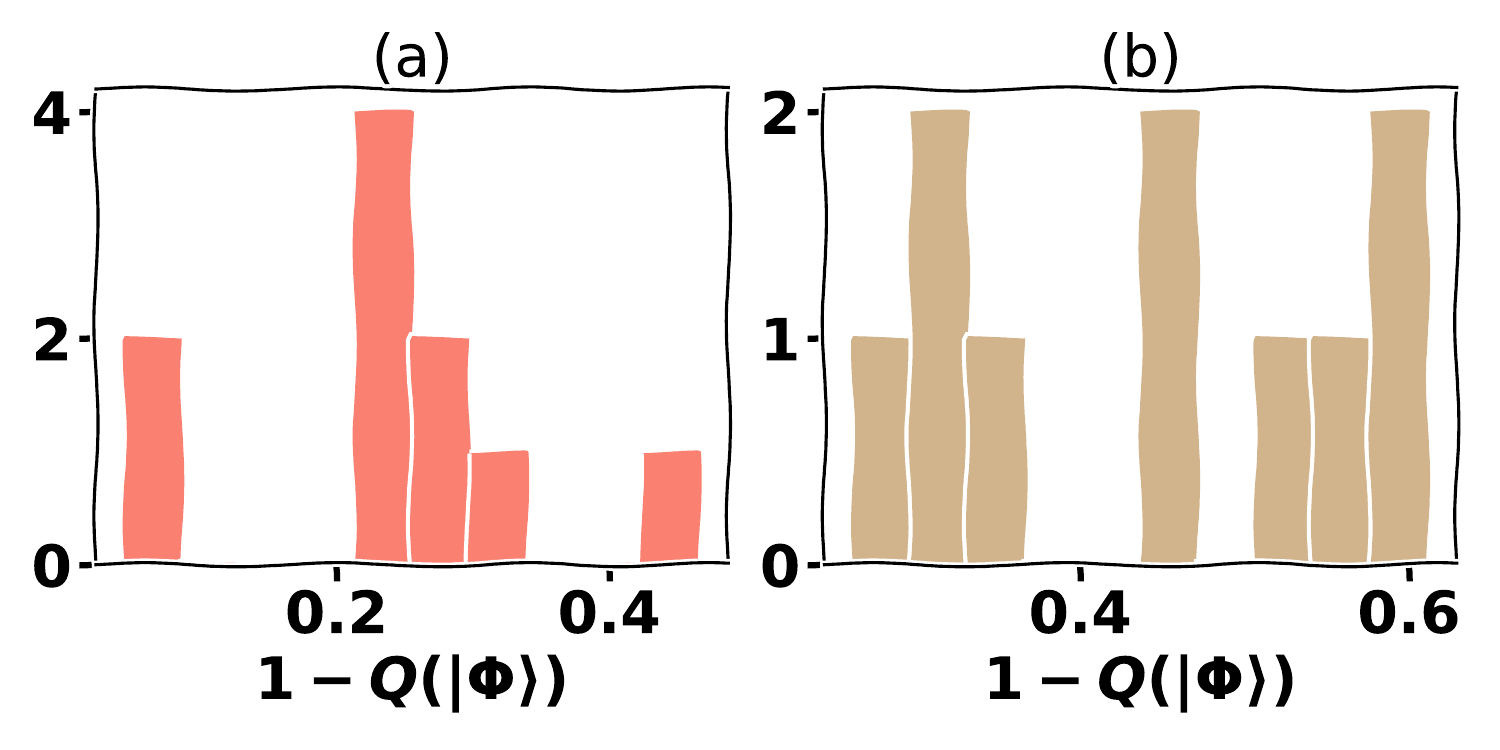}

    \caption{\justifying Histograms of 10 random topologies with amplitude noise on each wire after the $\mathrm{SWAP}$ gate that simulate the $\mathbf{U}^{\varkappa}$. Left histogram, $\mathcal{U}^{BM}_{\rm RN}$. right histogram, the $\mathcal{U}^{\theta}_{\rm RN}$ random topologies are compared. Also, for moderate noise, the better performances obtained from a nonlinear function are evident with minimal sampling.} 
    \label{fig:noisy 5 qubits}
\end{figure}
Even though the number of topologies here sampled is pretty low, we can immediately find one that can obtain entanglement values of $0.07$ and $0.31$ for the memristor type, and $0.09$ and $0.16$ for the sine nonlinear element. The same outcomes are also confirmed for SC+1 in Fig.\ref{fig:noisysin montecarlo}.

In the second histogram, we display the result obtained for the architecture $\mathcal{U}^{\rm sin}_{\rm RN}$ topology, compared against its $\mathcal{U}^{\rm \theta}_{\rm RN}$ counterpart. 

We can notice again that the hypothesis that the sine nonlinear components offer an evident advantage, and undoubtedly, for the global entanglement generation problem, this family of nonlinearity bestows high efficiency. Compared to Fig.\ref{fig:noisy 5 qubits} we can appreciate that both $\uscone^{\rm BM}$ and $\uscone^{\rm sin}$ are comparable.

\section{10 qubits with amplitude damping noise}
\label{Sec:10 noisy qubits}
In this section, we provide the outcomes obtained from the MW optimization for 10 qubits with $\mathcal{U}^{BM(t)}$ and $\mathcal{U}^{\rm sin(t)}$, and $\mathcal{T}_{\rm ladder}$. We can see that the topology close to the linear graph state is extremely efficient and can achieve a value of MW close to a GHZ.
\begin{figure}[h!]
    \centering
    \includegraphics[width=0.8\linewidth]{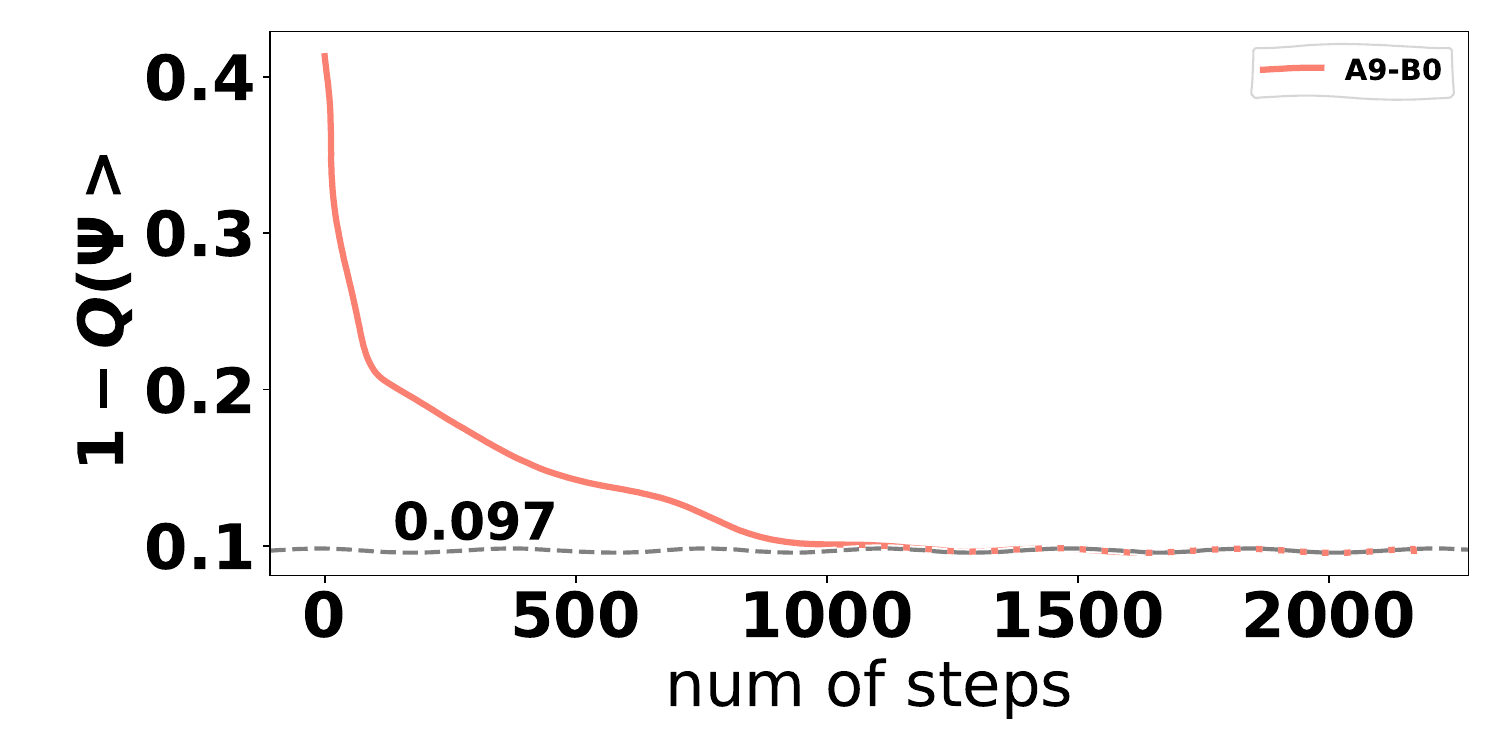}
    \includegraphics[width = 0.8\linewidth]{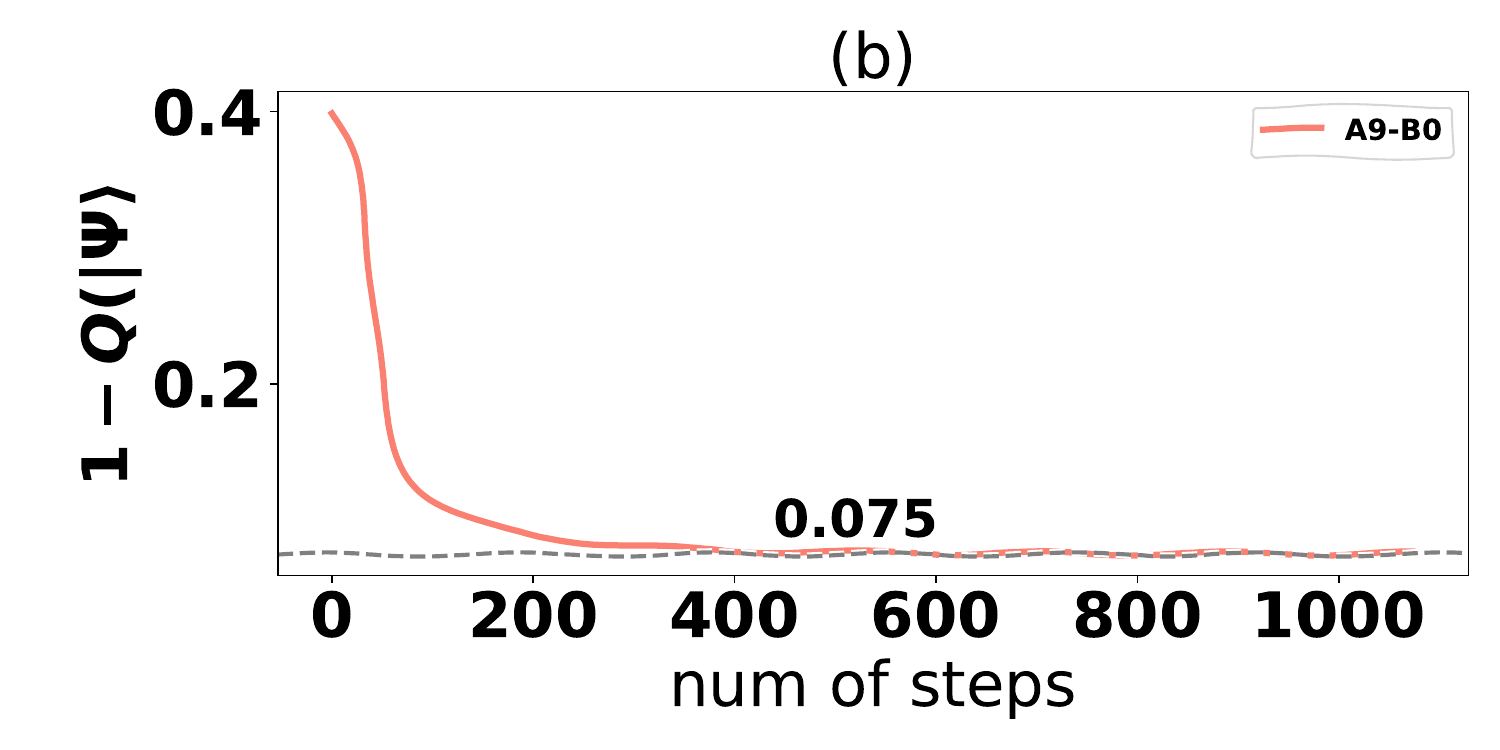}
    \caption{\justifying Panel (a). The 10 qubits $\uscone^{\rm BM}$ model for amplitude damping only with $\gamma= 0.01$. The mild amount of noise does not hamper the global entanglement generation. Panel(b) the $\uscone^{\rm sin}$ ladder model. We can notice that the $sine$ nonlinear function provides faster convergence to the minima.  }
    \label{fig:10 qubits noisy}
\end{figure}

In this numerical experiment, we can see the entanglement engineering ability obtained from a 10-qubit network of $\uscone^{BM}$. The network oscillates around the optimal minima, and obtains a value of $\sim 0.98$ of MW entanglement, and $\sim 0.75$ for the $\uscone^{\rm BM}$ and $\uscone^{\rm sin}$ model,s respectively. The memristor reflectivity enhanced a greater stability inside the training, enabling the early-stopper to operate and stop the simulation before reaching 2000 epochs. 
\newpage
\section{20 qubits experiment}
\label{sec:20 qubits noiseless}

\begin{figure}[h!]
    \centering
    \includegraphics[width=0.8\linewidth]{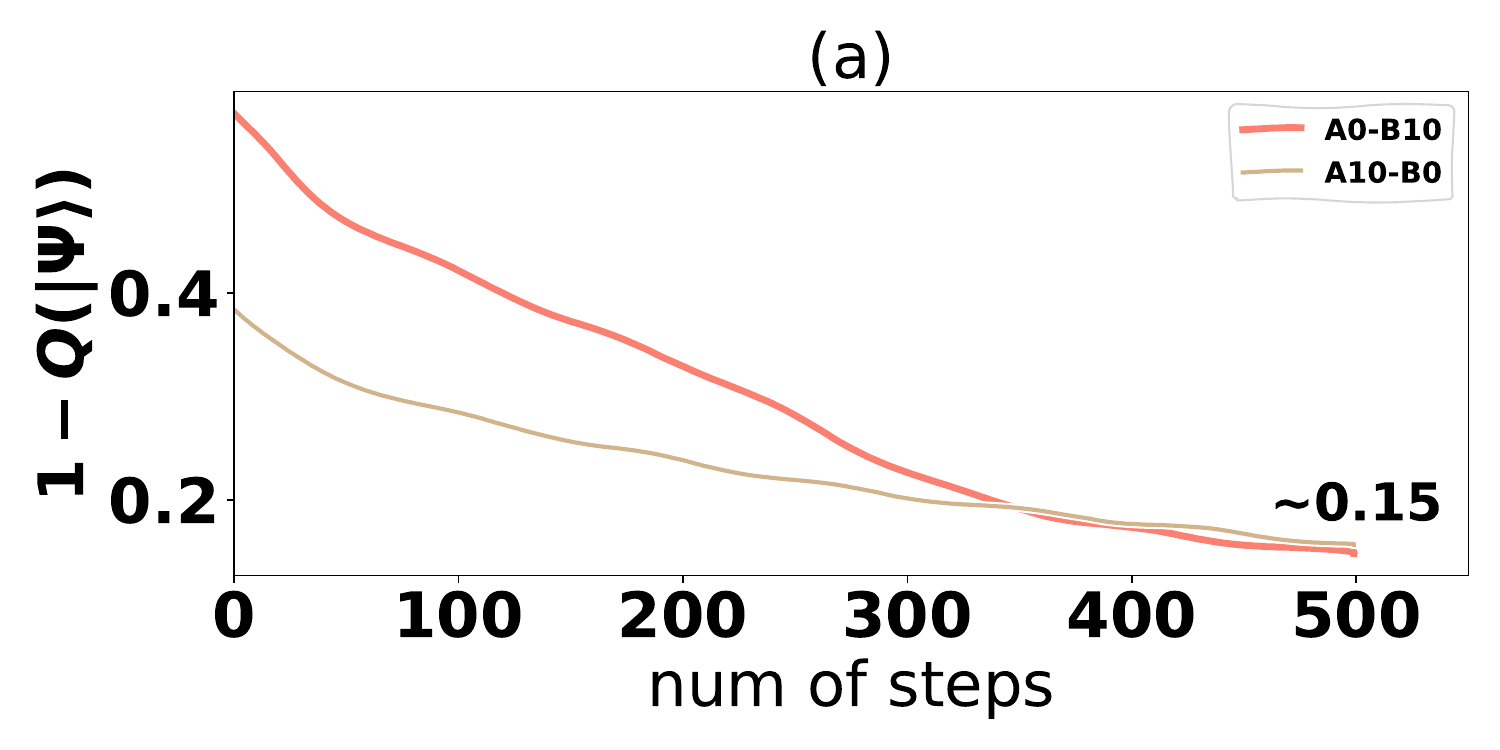}
    \includegraphics[width=0.8\linewidth]{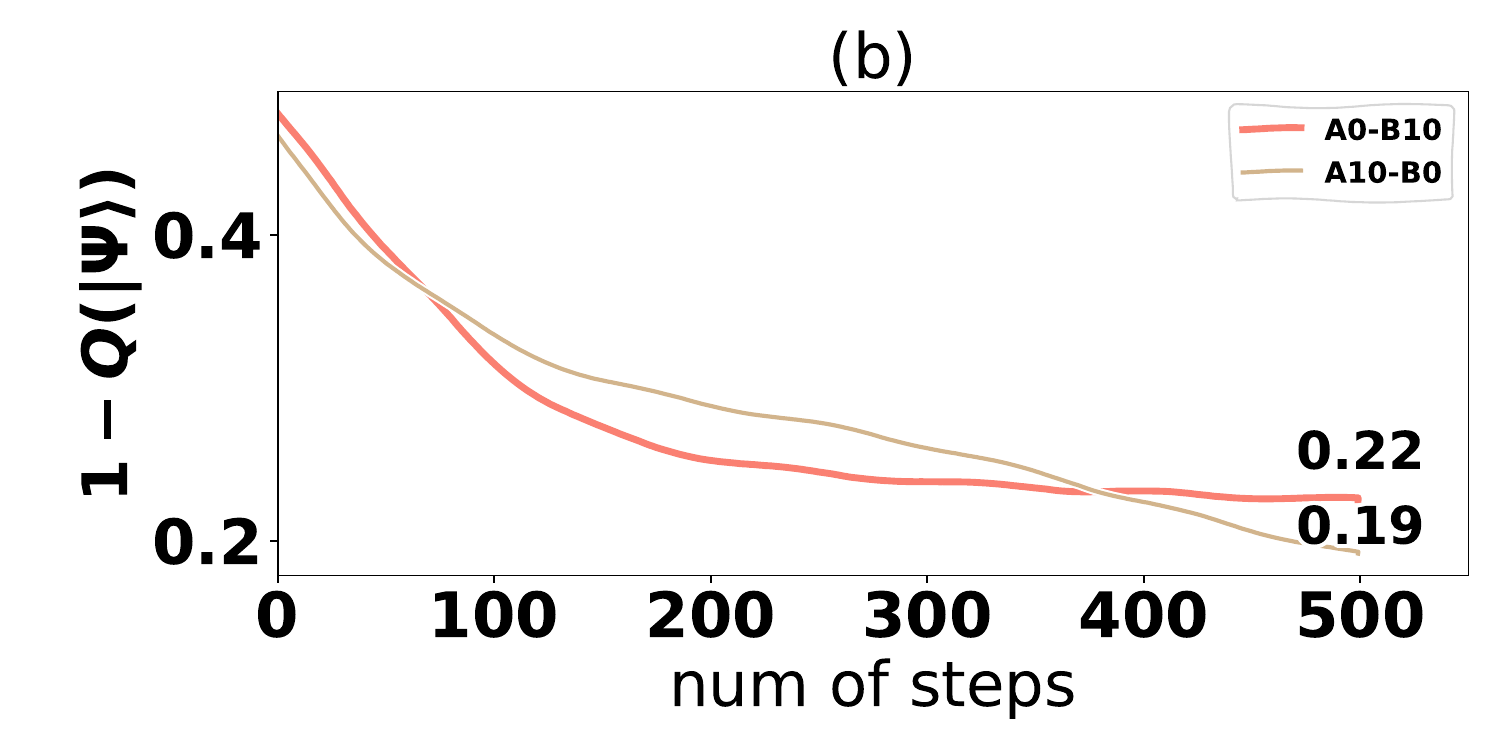}
    \caption{\justifying 20 qubit optimization. Upper panel, quantum network optimization with memristor-inspired reflectivity term. We can see that the topologies are equivalent. Lower panel, sinusoidal nonlinear term application. We can check that the topology with the control term on the last wire (the number zero) is bound to achieve a higher amount of entanglement.   }
    \label{fig:20 qubits}
\end{figure}

An open question is how the scaling in the number of qubits may affect the outcomes. 
In Fig.\ref{fig:20 qubits} we compare the beam splitter nonlinear function (upper panel) against the simple sinusoidal (lower panel). 
This simulation is provided for a very low amount of noise, with only amplitude damping after the two $\mathrm{SWAP}s$ for a $\gamma= 10^{-3}$. 

For almost pure states, the memristor-inspired term in Eq.~\eqref{eq: reflectivity} offers better performance in general. We have already detailed in Sec.\ref{sec:methods} that this does not hold anymore for the realistic noisy case, where now the differential panorama is completely modified by the presence of local noise.

\section*{Appendix: Technical Derivation and Interpretation of the Response Function}

In recent years, a revival of photonic applications to extracting quantum state properties (or features) is showing a resurgence, with several works on photonic platforms \cite{ExperimentalPropertiesPhotonicPlatform, ReviewReservoir-Innocenti2023, Reservoir-Spagnolo2022}.
\section{The beam splitter reflectivity}

In this work, we initiate our analysis by adopting a circuit-based model of a photonic quantum memristor, as presented in the recent study~\cite{AlbertoMemristor}. 
The response function
\begin{equation}
R(t) = \frac{T_{\mathrm{osc}}}{T_{\mathrm{int}}\,4\pi}
\left[
\sin\!\left(\frac{2\pi t - 2\pi T_{\mathrm{int}}}{T_{\mathrm{osc}}}\right)
-
\sin \left(\frac{2\pi t}{T_{\rm osc}}\right)
\right],
\label{eq:response function}
\end{equation}
encapsulates the temporal behaviour of an oscillatory system whose dynamics are influenced by a finite integration window $T_{\rm int}$ and an intrinsic oscillation period $T_{\rm osc}$. 
The term 
$\sin\,\left[(2\pi t - 2\pi T_{\rm int})/T_{\rm osc} \right]$
corresponds to the oscillation evaluated one integration interval prior, thereby encoding the system’s memory of past dynamics, whereas 
$\sin\,\Big[(2\pi t)/T_{\rm osc} \Big]$
represents the instantaneous, undelayed oscillatory component. 
Accordingly, the function $R(t)$ serves as a precise indicator of both residual coherence and the temporal modulation imparted by memristive dynamics, embodying the hallmark history-dependent behaviour of memory-enabled quantum systems.

\section{Structure of the Response Function}

The response function considered in the main text,
\begin{equation}
R(t) = \frac{T_{\mathrm{osc}}}{T_{\mathrm{int}}\,4\pi}
\left[
\sin\!\left(\frac{2\pi\,t - 2\pi\,T_{\mathrm{int}}}{T_{\mathrm{osc}}}\right)
-
\sin\!\left(\frac{2\pi\,t}{T_{\mathrm{osc}}}\right)
\right],
\label{eq:R_appendix}
\end{equation}
emerges from the evaluation of an oscillatory process over a finite memory window of temporal width $T_{\mathrm{int}}$. 
This structure reflects the difference between two phase-shifted sinusoidal components, each normalized by an amplitude factor depending on both the oscillation period and the effective integration interval. 

To understand Eq.~\eqref{eq:R_appendix}, we begin by noting that an ideal periodic signal of angular frequency 
\[
\omega = \frac{2\pi}{T_{\mathrm{osc}}},
\]
is integrated over the shifted interval $[t - T_{\mathrm{int}},\, t]$.  
For an oscillatory function of the form $f(\tau) = \sin(\omega \tau)$, the finite-time integral can be computed exactly:
\begin{equation}
\int_{t - T_{\mathrm{int}}}^{t} \sin(\omega \tau)\, d\tau
= \frac{1}{\omega} 
\left[
\cos\!\left(\omega(t - T_{\mathrm{int}})\right)
-
\cos(\omega t)
\right].
\label{eq:integral_exact}
\end{equation}

Using trigonometric identities and normalizing by $T_{\mathrm{int}}$, one obtains a difference of sines whose argument is shifted by $T_{\mathrm{int}}$, leading to the structure displayed in Eq.~\eqref{eq:R_appendix}.  
The prefactor $\frac{1}{\omega}$ arises naturally from the integral of the sinusoidal function and provides the correct amplitude scaling.

\subsection*{Physical Interpretation}

In systems exhibiting memory-dependent (or ``memristive'') behavior, the present response does not depend solely on the instantaneous value of the driving field, but also on its history within a specific temporal window. 
Eq.~\eqref{eq:R_appendix} captures exactly this effect: $R(t)$ measures the deviation introduced by integrating the oscillatory signal over an interval $T_{\mathrm{int}}$, effectively encoding a temporal lag between the free oscillation and its memory-processed counterpart.

The two sinusoidal terms in $R(t)$ may be interpreted as:
\begin{align}
    \sin\!\left[(2\pi t)/T_{\mathrm{osc}}\right]
    \sin\!\left[(2\pi t - 2\pi T_{\mathrm{int}})/T_{\mathrm{osc}}\right], 
\end{align}
With the first equation, the instantaneous oscillation of the underlying coherent process, and the second equation the the oscillation evaluated one integration window earlier.
The difference between these terms provides a direct measure of the accumulated phase displacement induced by memory effects. 
In oscillatory regimes where $T_{\mathrm{int}} \ll T_{\mathrm{osc}}$, this phase shift is small, and $R(t)$ behaves nearly linearly in $T_{\mathrm{int}}$.  
Conversely, when $T_{\mathrm{int}}$ becomes comparable to $T_{\mathrm{osc}}$, the interference between the two sinusoidal contributions becomes pronounced, generating richer dynamical features in $R(t)$.

This formulation offers a compact yet powerful tool to quantify the interplay between coherence, periodic driving, and memory depth.  
In the context of photonic or quantum memristive systems, the response function Eq.~\eqref{eq:R_appendix} is particularly useful for characterizing the temporal retention of coherence and for distinguishing between purely dissipative and genuinely history-dependent dynamics. 
\newpage
\section{PPT-mixture SDP criterion for certifying GME.}
\label{app:SDP}

Let $\rho$ be an $n$-qubit density operator. A mixed state is \emph{biseparable} if it can be written as a convex combination of states that are separable with respect to (possibly different) bipartitions, i.e.,
\begin{eqnarray}
\rho &\in& \mathrm{BISep}
\iff
\rho=\sum_k p_k\,\rho_k,\\
\quad p_k&\ge& 0,\ \sum_k p_k=1, \nonumber\\
\rho_k\in &\mathrm{Sep}&\!\left(A_k\middle|A_k^c\right).\nonumber
\end{eqnarray}
A state is \emph{genuinely multipartite entangled} (GME) if $\rho\notin \mathrm{BISep}$. Since separability across a cut $A|A^c$ implies positivity under partial transposition (PPT) across that cut, $\rho^{T_A}\succeq 0$, the set of \emph{PPT mixtures} provides an efficiently testable outer approximation $\mathrm{BISep}\subseteq \mathrm{PPTmix}$. Given a collection $\mathcal{P}$ of nontrivial bipartitions, the PPT-mixture feasibility SDP reads: find operators $\{\sigma_A\}_{A\in\mathcal{P}}$ such that
\begin{equation}
\rho = \sum_{A\in\mathcal{P}} \sigma_A,\qquad
\sigma_A \succeq 0,\qquad
\sigma_A^{T_A}\succeq 0\ \ \forall A\in\mathcal{P},
\end{equation}

where $T_A$ denotes partial transposition on subsystem $A$ in the computational basis. If this SDP is \emph{infeasible}, then $\rho\notin \mathrm{PPTmix}$ and, because $\mathrm{BISep}\subseteq \mathrm{PPTmix}$, $\rho$ is certified to be \emph{GME}. If the SDP is feasible, the test is inconclusive (the state may still be GME). The dual program yields a Hermitian entanglement witness $W$ that separates $\rho$ from $\mathrm{PPTmix}$, i.e.\ $\mathrm{Tr}(W\sigma)\ge 0$ for all $\sigma\in \mathrm{PPTmix}$ while $\mathrm{Tr}(W\rho)<0$.

\end{document}